\documentclass[12pt,preprint]{elsarticle}

\usepackage{url}
\usepackage{rotating}
\usepackage{booktabs}
\usepackage{eqnarray}
\usepackage{amsmath}
\usepackage{tabularx}
\usepackage{enumitem} 
\usepackage{tablefootnote} 
\usepackage{subfig} 
\usepackage{dcolumn} 

\journal{Journal of Quantitative Spectroscopy and Radiative Transfer }
\begin{document}

\begin{frontmatter}

\title{Recovering the city street lighting fraction from skyglow measurements in a large-scale municipal dimming experiment}

\author[ida,cdss]{John C.~Barentine\corref{cor1}}
\ead{john@darksky.org}

\author[comenius]{Franti\v{s}ek Kundracik}
\ead{frantisek.kundracik@fmph.uniba.sk}

\author[comenius,ica]{Miroslav Kocifaj}
\ead{Miroslav.Kocifaj@savba.sk}

\author[cot]{Jessie C.~Sanders}
\ead{Jessie.Sanders@tucsonaz.gov}

\author[flwo,psi]{Gilbert A.~Esquerdo}
\ead{esquerdo@psi.edu}

\author[ida,cdss]{Adam M.~Dalton}
\ead{adam@darksky.org}

\author[ida,cdss]{Bettymaya Foott}
\ead{bettymaya@darksky.org}

\author[catalina]{Albert Grauer}
\ead{algrauer@me.com}

\author[starizona]{Scott Tucker}
\ead{scott@starizona.com}

\author[gfz,igb]{Christopher C.~M.~Kyba}
\ead{kyba@gfz-potsdam.de}

 \cortext[cor1]{Corresponding author}

\address[ida]{International Dark-Sky Association, 3223 N.~1st Ave, Tucson, AZ 85719 USA}
\address[cdss]{Consortium for Dark Sky Studies, University of Utah, 375 S 1530 E, RM 235 ARCH, Salt Lake City, Utah 84112-0730 USA}
\address[comenius]{Faculty of Mathematics, Physics, and Informatics, Comenius University, Mlynsk\'{a} Dolina, 842 48 Bratislava, Slovakia}
\address[ica]{ICA, Slovak Academy of Sciences, D\'{u}bravsk\'{a} Road 9, 845 03 Bratislava, Slovakia}
\address[cot]{City of Tucson, 255 W.~Alameda St, Tucson, AZ 85701 USA}
\address[starizona]{Starizona, 5757 North Oracle Rd, Tucson, AZ 85704 USA}
\address[flwo]{Fred Lawrence Whipple Observatory, P.O. Box 6369, Amado, AZ 85645-6369 USA}
\address[psi]{Planetary Science Institute, 1700 East Fort Lowell, Suite 106, Tucson, AZ 85719-2395 USA}
\address[catalina]{Catalina Sky Survey, Department of Planetary Sciences, Lunar and Planetary Laboratory, 1629 E.~University Blvd., Tucson, AZ 85721-0092 USA}
\address[gfz]{GFZ German Research Centre for Geosciences, Potsdam 14473, Germany} 
\address[igb]{Leibniz Institute of Freshwater Ecology and Inland Fisheries (IGB), Berlin 12587, Germany}

%
%
 \begin{abstract}
Anthropogenic skyglow dominates views of the natural night sky in most urban settings, and the associated emission of artificial light at night (ALAN) into the environment of cities involves a number of known and suspected negative externalities. One approach to lowering consumption of ALAN in cities is dimming or extinguishing publicly owned outdoor lighting during overnight hours; however, there are few reports in the literature about the efficacy of these programs. Here we report the results of one of the largest municipal lighting dimming experiments to date, involving $\sim$20,000 roadway luminaires owned and operated by the City of Tucson, Arizona, U.S. We analyzed both single-channel and spatially resolved ground-based measurements of broadband night sky radiance obtained during the tests, determining that the zenith sky brightness during the tests decreased by ($-5.4\pm0.9$)\% near the city center and ($-3.6\pm0.9$)\% at an adjacent suburban location on nights when the output of the street lighting system was dimmed from 90\% of its full power draw to 30\% after local midnight. Modeling these changes with a radiative transfer code yields results suggesting that street lights account for about ($14\pm1$)\% of light emissions resulting in skyglow seen over the city. A separate derivation from first principles implies that street lighting contributes only $2-3$\% of light seen at the zenith over Tucson. We discuss this inconsistency and suggest routes for future work.
\end{abstract}
 

\end{frontmatter}

%
%
\section{Introduction}

Light pollution is a global phenomenon caused by the prolific use of artificial light at night (ALAN).~\citep{Falchi2016,Kyba2017} ALAN offers clear benefits to human society by ensuring safe transit at night, enabling the nighttime economy, and enhancing public perception of outdoor spaces at night through placemaking,~\citep{Boyce2019} but its use entails a number of known and suspected hazards to the natural nocturnal environment, e.g.,~\citep{Minnaar2015,Bennie2016,Bennie2018}, including potentially significant disruption of ecosystem services~\citep{Lyytimaki2013,Knop2017,Grubisic2018} and threats to biodiversity.~\citep{Guette2018,Koen2018} On the other hand, actively preserving natural nighttime darkness not only appears to convey environmental benefits, but also can support sustainable rural economic development through `astrotourism.'~\citep{Collison2013,Labuda2015,Labuda2016}

Given the negative environmental influence of ALAN, activists have called for addressing the problem through various lighting engineering and public policy means. Achieving these goals requires identifying best practices for reducing light emissions to levels strictly necessary to ensure public safety. The efficacy of these practices is in part determined by the participation rate of owners of light sources: the more sources whose output is reduced or eliminated presumably results in a greater reduction in the amount of ALAN in the nighttime environment. Public lighting is an attractive target for reduction efforts, especially in urban settings, given that many thousands of luminaires are often under the control of a single administrative entity. 

Limited evidence exists to date suggesting that modifications to existing municipal lighting systems, in particular, can yield measurable environmental effects.~\citep{Aube2014,Barentine2018} However, current models are acutely deficient in the sense that the fraction of total light emissions comprised by municipal lighting in cities is not well constrained. While some cities have created inventories of publicly owned lighting, including light emission parameters of luminaires, a complete accounting of privately owned lighting is available only for the smallest municipalities. Currently available remote sensing data do not have sufficient spatial resolution to reliably determine the ownership status of lighting, although one approach to solving that problem is to make an initial guess of the function of outdoor lighting by matching the distribution of remotely sensed night lights to maps of land use patterns.~\citep{Kuechly2012} However, even high-resolution aerial imagery of cities is not presently subject to analysis techniques that can reliably discern between lighting types. 

In order to better inform models, there is a need to determine the relative contributions of public and private sources to the total light emissions of cities. Reducing the influence of light pollution in cities may be well served by focusing on public lighting; however, if public lighting were a small fraction of total lighting, then the impact of such efforts would be proportionately less. Determining the fraction of total lighting attributable to publicly owned sources in cities is therefore important to guide decisions on how best to direct mitigation efforts. 

Previous attempts to determine the relative contributions of public and private lighting in cities have found public lighting fractions ranging from $\sim10-75$\% of total urban light emissions.~\citep{Barentine2018,Kuechly2012,Luginbuhl2009,Hiscocks2010,Bara2018}. Several methods have been employed to determine these values, most of which rely on certain model assumptions. Perhaps the most robust approach to distinguish public lighting from private lighting is to change the output of only public light sources in some known way and measure by how much the total city light emission changes. This is relatively easy to arrange for smaller villages (e.g.,~\citep{Jechow2018}), but it is difficult for large cities. We arranged a test with the municipal government of Tucson, Arizona, U.S., that involved dimming the municipal street lighting system by certain amounts on a series of test nights. By controlling the output of street lights, we recovered the fraction of the total light emission of Tucson represented by the street lights through a combination of measurements of the brightness of the night sky and modeling skyglow with a radiative transfer code.

This paper is organized as follows. First, in Section~\ref{sec:test}, we outline the parameters of the street lighting dimming experiment and the intended goals of the project. Next, in Section~\ref{sec:measurements}, we describe the measurements made during each of the test nights. Then, in Section~\ref{sec:simulations}, we present the results of radiative transfer model runs used to predict skyglow changes over the city consequent to the dimming tests. We analyze the observations in the context of the model results in Section~\ref{sec:analysis}, and finally we summarize our work, point out its limitations, and offer suggestions to guide future experiments of this nature in Section~\ref{sec:summary}.

%
%
\section{Lighting System and Test Parameters}\label{sec:test}

\begin{figure}[htp]
\centering
\includegraphics[width=\textwidth]{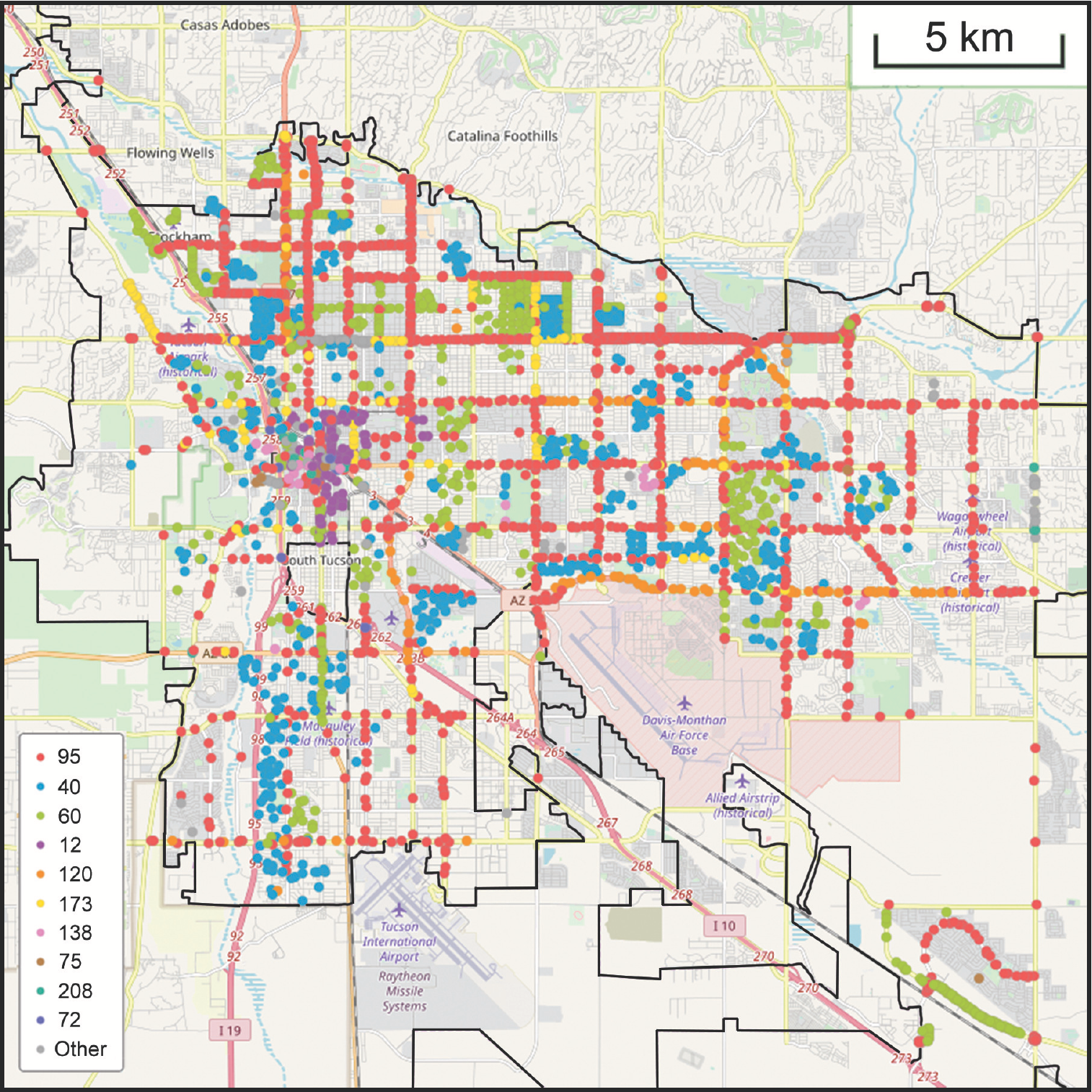}
\caption{A map of the greater Tucson, Arizona, metropolitan area on which is overlaid markers representing the locations of individual luminaires comprising the Tucson municipal street lighting system. The power draw of each luminaire in Watts is indicated by the legend at lower left, whose symbols are ordered according to the prevalence of luminaire Wattages. The map is oriented to the usual cartographic convention with north at top and east at right, and a 5-kilometer scale bar is provided in the upper-right corner. The legal boundaries of the City of Tucson correspond to the thin black lines; note that only the city regions in which known street lighting is present are included in this map. Lighting information is provided courtesy of the City of Tucson Open Data portal; base map copyright OpenStreetMap contributors and licensed under CC BY-SA.}
\label{Tucson-streetlight-map}
\end{figure}

Tucson is a city of 535,000 inhabitants; the population of its metropolitan statistical area (MSA) is about one million. The Tucson municipality owns and manages approximately 20,000 roadway luminaires distributed across 587 square kilometers of its incorporated territory (Figure~\ref{Tucson-streetlight-map}); roadway luminaires operated by other municipalities in the MSA did not take part in the tests reported here. The inventory of luminaires consists of over 90\% white LED products with a nominal correlated color temperature of 3000 kelvins and the remainder composed of legacy technologies (specifically, a mixture of high- and low-pressure sodium lamps).~\citep{Barentine2018}  Of the LED luminaires, 19,561 are network-addressable and comprise a total light emission of $\sim$2.0$\times$10$^{8}$ lumens when operated at their full rated electric power draw.\footnote{Lighting data are publicly available from the City of Tucson in Open GIS format on \url{https://hub.arcgis.com/datasets/09ed59b6aae2483aa1bd32837d4aa7e5_19}.} 

To prolong the service lifetimes of the luminaires and adjust for expected lumen depreciation during field operations, the system is operated routinely at 90 percent of full power draw, for a nominal total light emission of $\sim$1.8$\times$10$^{8}$ lumens under normal conditions. Since shortly after its installation in 2016--17, the LED street lighting system has been subjected to a nightly dimming scheme in which the electric power draw of $\sim$16,000 luminaires, or $\sim$83\% of all luminaires in the system, are reduced to 60 percent of full power at midnight; they remain in this configuration until they are routinely extinguished 30 minutes after local sunrise. The set of luminaires dimmed in this fashion are located throughout the city. The nominal total light emission of the system in the 60 percent dimmed configuration is $\sim$1.2$\times$10$^{8}$ lumens.

Tests of the municipal street lighting system were conducted by the City of Tucson on the nights of UTC 29 March through 8 April 2019. The tests involved altering the nightly dimming program of the network-addressable LED luminaires, avoiding those located at road intersections, pedestrian crosswalks, and areas of high pedestrian activity occurring post-midnight due to public safety concerns. Luminaires in the test subject to routine nightly dimming were further reduced to 30 percent of their full power draw on the nights of UTC 29 March through UTC 3 April.\footnote{To avoid confusion, Coordinated Universal Time (UTC) dates are used exclusively here to specify test nights such that a change in local calendar date did not occur during a test night.} On the following five nights (UTC 4-8 April), the same set of luminaires was programmed to brighten to 100 percent of full power at midnight. Each afternoon, municipal engineers sent the dimming programs to the test luminaires. The following morning, they received a report from the lighting control system specifying, for each luminaire, a unique identification number, geoposition, network address, nominal full power draw, and the low and high power draw values achieved during the night. From this information we determined the light output of each luminaire during each test night. The light output of the each luminaire before and after midnight, combined with the position information, yields a precise model of the component of the CEF specifically attributable to the fraction of the municipal lighting system under active control. 

A summary of the lighting reports is shown in Table~\ref{test-percentages}, indicating the number of luminaires in the test set that achieved various minimum power draws from zero to 100 percent in increments of 10 percent of full power. On the test nights, about 81 percent of luminaires in the test set successfully responded to the dimming program and reported achieving the commanded minimum power draw at midnight. An important exception is the night of UTC 4 April, on which most luminaires executed the previous night's program. We intended to observe several adjacent nights, during which the usual dimming program was executed, as controls but poor weather precluded making valid sky brightness measurements.

\begin{sidewaystable}
\medskip
\resizebox{\linewidth}{!}{%
\begin{tabular}{ lccccccccccccc }
 \toprule
 \textbf{} & \textbf{} & \textbf{} & \textbf{} & \textbf{} & \textbf{} & \textbf{} & \textbf{} & \textbf{} & \textbf{} & \textbf{} & \textbf{} & \textbf{Percent} \\
 \textbf{UTC Date} & \textbf{0\%} & \textbf{10\%} & \textbf{20\%} & \textbf{30\%} & \textbf{40\%} & \textbf{50\%} & \textbf{60\%} & \textbf{70\%} & \textbf{80\%} & \textbf{90\%} & \textbf{100\%} & \textbf{Successful} \\
 \hline
28 Mar	&	1439	&	70	&	6	&	15	&	85	&	192	&	14851	&	98	&	540	&	1835	&	415	&	91.2	\\
29 Mar	&	1411	&	101	&	98	&	11872	&	208	&	112	&	2971	&	21	&	355	&	1989	&	408	&	73.5	\\
30 Mar	&	1313	&	75	&	96	&	12863	&	219	&	66	&	2107	&	22	&	319	&	2039	&	425	&	79.4	\\
31 Mar	&	1235	&	81	&	101	&	13345	&	227	&	58	&	1693	&	17	&	632	&	1725	&	431	&	80.4	\\
01 Apr	&	1209	&	93	&	107	&	13809	&	272	&	57	&	1186	&	14	&	588	&	1789	&	419	&	83.4	\\
02 Apr	&	1218	&	60	&	102	&	13966	&	248	&	56	&	1093	&	14	&	438	&	1927	&	420	&	85.1	\\
03 Apr	&	1186	&	84	&	100	&	13984	&	221	&	65	&	1092	&	19	&	851	&	1513	&	426	&	2.5	\\
04 Apr	&	1161	&	51	&	29	&	2725	&	76	&	94	&	1049	&	55	&	153	&	3147	&	10863	&	71.2	\\
05 Apr	&	1176	&	65	&	10	&	794	&	28	&	69	&	830	&	60	&	125	&	697	&	15509	&	87.7	\\
06 Apr	&	1164	&	38	&	16	&	628	&	31	&	61	&	775	&	68	&	87	&	537	&	15949	&	89.3	\\
07 Apr	&	1128	&	41	&	8	&	502	&	41	&	62	&	801	&	83	&	118	&	7952	&	8679	&	82.8	\\
 \bottomrule
\end{tabular}}
\caption{As-reported dimming schedule for City of Tucson-owned roadway luminaires during the March-April 2019 test showing the number of luminaires achieving the stated percentages of full power. Luminaires reporting $>$100\% nominal full power consumption are not included; these comprised at most 236 luminaires, or $\sim$1.2\% of known lights, on any given test night, a fraction too small to affect our analysis. Most luminaires reporting 0\% did not communicate any information, while some did not light; it is impossible to distinguish among these in the available data. The last column indicates the percent of luminaires in the test set that successfully achieved the commanded dimming state each night.}
\label{test-percentages}
\end{sidewaystable}

%
%
\section{Skyglow Measurements}\label{sec:measurements}

We made a series of measurements of the brightness of the night sky over and adjacent to Tucson during the dimming tests in order to quantify the degree to which the sky brightness changed as a result of the different street light dimming configurations. 

\subsection{Data sources and acquisition}

\begin{figure}[htp]
\centering
\includegraphics[width=\textwidth]{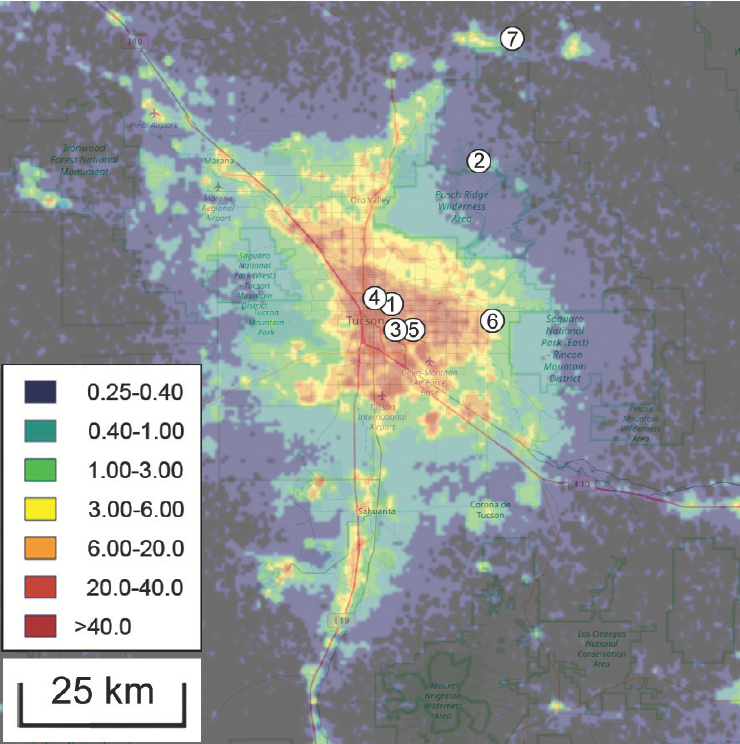}
\caption{A map of the greater Tucson, Arizona, metropolitan area on which is overlaid a set of false colors representing upward-directed broadband radiances detected by the Visible Infrared Imaging Radiometer Suite Day-Night Band (VIIRS-DNB) aboard the U.S. National Oceanic and Atmospheric Administration \emph{Suomi NPP} satellite in March 2019. The legend at lower left gives the values of the colors in radiance ranges of 0.25 (darkest blue) to 40 (red) in units of nW cm$^{-2}$ sr$^{-1}$. The white markers, numbered 1 to 8, correspond to the sky brightness measurement locations listed in Table~\ref{equipment}. The map is oriented to the usual cartographic convention with north at top and east at right, and a 25-kilometer scale bar is provided in the lower-left corner. Base map copyright OpenStreetMap contributors and licensed under CC BY-SA; VIIRS-DNB radiance data courtesy Jurij Stare (lightpollutionmap.info).}
\label{Tucson-data-source-map}
\end{figure}

We used a variety of broadband instruments, summarized in Table~\ref{equipment}, to directly detect the radiance of the night sky during the tests. The locations of the sensors during the observations is shown in Figure~\ref{Tucson-data-source-map}. 

\begin{sidewaystable}
\medskip
\resizebox{\linewidth}{!}{%
\begin{tabular}{ llccccccc }
 \toprule
  \textbf{} & \textbf{} & \textbf{} & \textbf{Serial ID or} & \textbf{Site}& \textbf{Latitude} & \textbf{Longitude} & \textbf{Elevation} & \textbf{$D$} \\
 \textbf{Source} & \textbf{Measurement} & \textbf{Reference} & \textbf{Description} & \textbf{Number\footnote{These numbers refer to the markers shown on the map in Figure~\ref{Tucson-streetlight-map}}} & \textbf{(deg)} & \textbf{(deg)} & \textbf{(m)} & \textbf{(km)\footnote{Radial distance from the Tucson city center (32.221878, -110.971271, 727 m).}} \\
 \hline
 SQM-L & Single-channel luminance, & \citep{Cinzano2007} & 4246 & 1 & 
 32.25086 & -110.93854 & 738 & 3.5 \\
  & narrow acceptance angle & & 4387 & 2 &  32.44266 & -110.78889 & 2787 & 29.0 \\
  & & & 5442 & 3 &  32.21239 & -110.92515 & 756 & 5.5 \\ 
  & & & 10473\footnote{Position on the night of UTC 31 March.} & 4 &  32.25204 & -110.94811 & 733 & 3.4 \\
  & & & 10473\footnote{Position on the night of UTC 1 April.} & 5 &  32.21350 & -110.90542 & 768 & 6.3 \\    
 TESS-W & Single-channel luminance & \citep{Zamorano2017,Bara2019} & stars19 & 6 &  32.22388 & -110.76741 & 835 & 20.0 \\
  & & & stars227 & 7 &  32.60985 & -110.73369 & 1316 & 49.0 \\
 DSLR camera + & Spatially-resolved, & \citep{Kollath2017} & Canon T2i DSLR + & 3 &  32.21239 & -110.92515 & 756 & 5.5 \\
 fisheye lens & multichannel luminance & & Sigma 4.5mm f/2.8 lens & & & \\
 \bottomrule
\end{tabular}}
\caption{Summary of data sources used to characterize visual night sky brightness in this study.}
\label{equipment}
\end{sidewaystable}

The photometric data sources included four narrow-angle Sky Quality Meter (``SQM-L'') devices \citep{Cinzano2007}, two Telescope Encoder and Sky Sensor (TESS-W) units \citep{Zamorano2017,Bara2019}, and luminance-calibrated all-sky digital imagery.~\citep{Kollath2017} The SQM-L units we used were serial numbers 4246, 4387, 5442, and 10473. Of these, three are the handheld variety, while number 4387 is a data-logging device (SQM-LU-DL) permanently installed at the Catalina Sky Survey 1.5-meter telescope atop Mount Lemmon, 29 kilometers northeast of the Tucson city center. The TESS-W units used were stars19 and stars227. The stars19 unit was installed on the eastern edge of Tucson in May 2017 and has obtained nightly data since, whereas stars227 was installed at Oracle State Park just before the dimming test in March 2019. Only measurements of the radiance at the zenith were taken obtained using the single-channel devices. The stars227 unit was included as a control due to its radial distance from Tucson (49 kilometers from the city center) and the light-screening effect of the intervening Santa Catalina mountain range. At such distances we did not expect a priori to detect any artificial light component at the zenith attributable to light emission from Tucson. For the handheld SQM measurements, we discarded the initial reading in each set, which tends to be systematically brighter or darker than others in a series. This is an issue with the device known to the manufacturer and thought to result from slight internal heating of the sensor when power is initially applied.~\citep{Tekatch2017} 

For the all-sky imagery, we used an off-the-shelf Canon T5i digital single-lens reflex (DSLR) camera body and a Sigma 4 mm circular fisheye lens, giving an apparent field of view of 180 degrees. For consistent orientation of the resulting frames, the camera was pointed at the zenith with the base of the camera body facing the northern horizon. The camera settings for all light frames were 30-second exposures at f/2.8 and ISO 1600. Dark frames were obtained with the same settings but with the lens cap on, which were later subtracted from light images to remove the contribution from thermal noise in the camera electronics. The image sequence was: dark, dark, light, \ldots, light, dark, dark. We allowed the camera body to equilibrate to the ambient air temperature before recording images. No flat fields or other calibration data were obtained.

The dimming tests were planned for March and April 2019 as these months tend to have among the highest number of clear nights per month in southern Arizona as well as relatively low atmospheric turbidity. The test period was centered roughly on the date of new Moon (UTC 5 April), ensuring that all observations were made under conditions of astronomical darkness. For measurement sites in and adjacent to the Tucson MSA, we assumed that the contribution of light at the zenith from astronomical sources other than the Moon was negligibly small compared to anthropogenic light pollution. For more distant sites, we considered the possibility that astronomical light influenced the measurements, but decided that it did not. From the latitude of Tucson ($\sim$32$^{\circ}$ N), the North Galactic Pole transits near the zenith around midnight during March and April, minimizing the effect of the integrated light of the Milky Way on sky brightness measurements. We expected the contribution of zodiacal light to be similarly small, as the anti-solar point along the ecliptic approached the zenith by no more than 30 degrees on any of the dimming test nights. 

The test nights are listed in Table~\ref{test-2019-quality} along with availability of data from the sources listed above. We also make an overall quality assessment of each night in terms of weather conditions at Tucson International Airport\footnote{32.1145, -110.9392, 793 m; METAR code: KTUS.} recorded nightly at 0653 UTC.

\begin{sidewaystable}
\medskip
\resizebox{\linewidth}{!}{%
\begin{tabular}{ lcccccccccc }
 \toprule
 \textbf{UTC Date } & \textbf{Dimming} & \textbf{Quality} & \textbf{} & \textbf{SQM-L} & \textbf{SQM-L} & \textbf{SQM-L} & \textbf{SQM-LU-DL} & \textbf{FLWO} & \textbf{TESS} & \textbf{TESS}\\
\textbf{(2019)} & \textbf{Program} & \textbf{Assessment} & \textbf{DSLR} & \textbf{5442} & \textbf{4246} & \textbf{10473} & \textbf{4387} & \textbf{Camera} & \textbf{stars19} & \textbf{stars227} \\
\hline
29 Mar & 30\% & GOOD & YES & no & no & no & YES & YES & YES & YES \\
30 Mar & 30\% & FAIR & YES & YES & no & no & YES & YES & YES & YES \\
31 Mar & 30\% & GOOD & no & no & no & YES & YES & YES & YES & YES \\
1 Apr & 30\% & GOOD & YES & YES & no & YES & YES & YES & YES & YES \\
2 Apr & 30\% & FAIR & YES & YES & no & no & YES & YES & no & no \\
3 Apr & 30\% & GOOD & YES & YES & no & no & YES & YES & YES & YES \\
4 Apr & 100\% & FAIR & YES & YES & no & no & YES & YES & YES & YES \\
5 Apr & 100\% & POOR & YES & YES & no & no & YES & YES & YES & YES \\
6 Apr & 100\% & POOR & YES & YES & no & no & YES & YES & YES & YES \\
7 Apr & 100\% & GOOD & YES & YES & YES & no & YES & YES & YES & YES \\
8 Apr & 100\% & GOOD & YES & YES & no & no & YES & YES & YES & YES \\
9 Apr & 60\% & FAIR & YES & YES & no & no & YES & YES & YES & YES \\
10 Apr & 60\% & POOR & no & no & no & no & YES & YES & YES & YES \\
\bottomrule
 \end{tabular}}
\caption{Summary of night quality and data availability during the 2019 dimming tests. For each date (column 1), the programmed minimum luminaire power draw during the test is listed (column 2) along with a weather assessment (column 3). Night sky brightness data availability from sources listed in the main text is given in columns 4-11. In column 3, the quality assessments are as follows: GOOD means completely clear skies throughout the observation period under conditions appropriate for all-sky absolute photometry; FAIR means the data are compromised by the presence of clouds during the observation period rendering them inadequate for absolute photometry; and POOR means data were not obtained due to heavy clouds or overcast conditions.}
\label{test-2019-quality}
\end{sidewaystable}

\subsection{All-sky image data reduction}

All-sky images were calibrated using the method and software described by Koll\'{a}th and D\"{o}m\'{e}ny.~\citep{Kollath2017} The routines read the camera RAW-formatted images, apply spatial distortion/vignetting and luminance corrections, and output several data products. These include a calibrated version of the input image in cd m$^{-2}$; a Mercator-projected version of this image; and predicted SQM-L values in both magnitudes per square arcsecond (mag arcsec$^{-2}$), the native, logarithmic measurement unit of the SQM-L, and cd m$^{-2}$. The predictions are based on photometry of the calibrated images within software apertures of equivalent fields of view. The photometry was tied to lab calibration of the camera and lens combination and not to spectrophotometric standard stars or other field calibrators. 

\subsubsection{Performance of measurement devices}
In order to characterize the intrinsic uncertainties in our measuring devices, we inter-compared several of them side by side during cloud-free evening twilight in Tucson on UTC 29 April 2019. We tested our calibrated DSLR camera, SQM-L devices (serial numbers 4246, 5442, 9846 and 10473), and TESS-W stars19 over a luminance range spanning 20--9 magnitudes per square arcsecond, or  1 mcd m$^{-2}$ to $\sim$30 cd m$^{-2}$. Simultaneous measurements among individual devices correlated well with all other devices. The SQM-L devices have an precision of about ten percent ($\pm$0.1 magnitudes per square arcsecond, or $\sim0.3$ mcd m$^{-2}$) when compared to any other device; TESS-W stars19 showed a similar accuracy over the same luminance range. The DSLR sensor saturated at luminances above $\sim32$ mcd m$^{-2}$, and responded to light nonlinearly above $\sim20$ mcd m$^{-2}$. At lower luminances, it tracked linearly with the other devices. Its precision was better than the other devices by about a factor of two ($\pm0.05$ magnitude per square arcsecond; $\sim$0.2 mcd m$^{-2}$).

We found an offset of $-0.42\pm0.01$ magnitudes per square arcsecond ($+1.6\pm0.1$ mcd m$^{-2}$) between the DSLR and SQM-L unit 5442, used to make simultaneous measurements with the DSLR at Reid Park during the dimming tests. This offset seems to be consistent over a large range of luminances and is therefore interpreted as a difference in instrumental zeropoints. While acknowledging this somewhat large offset, we contend that it does not affect the relative brightness changes reported here.

\subsection{Photometry results}\label{subsec:phot-res}

\begin{figure}[htp]
\centering
\includegraphics[width=0.8\textwidth,trim=4 4 4 4,clip]{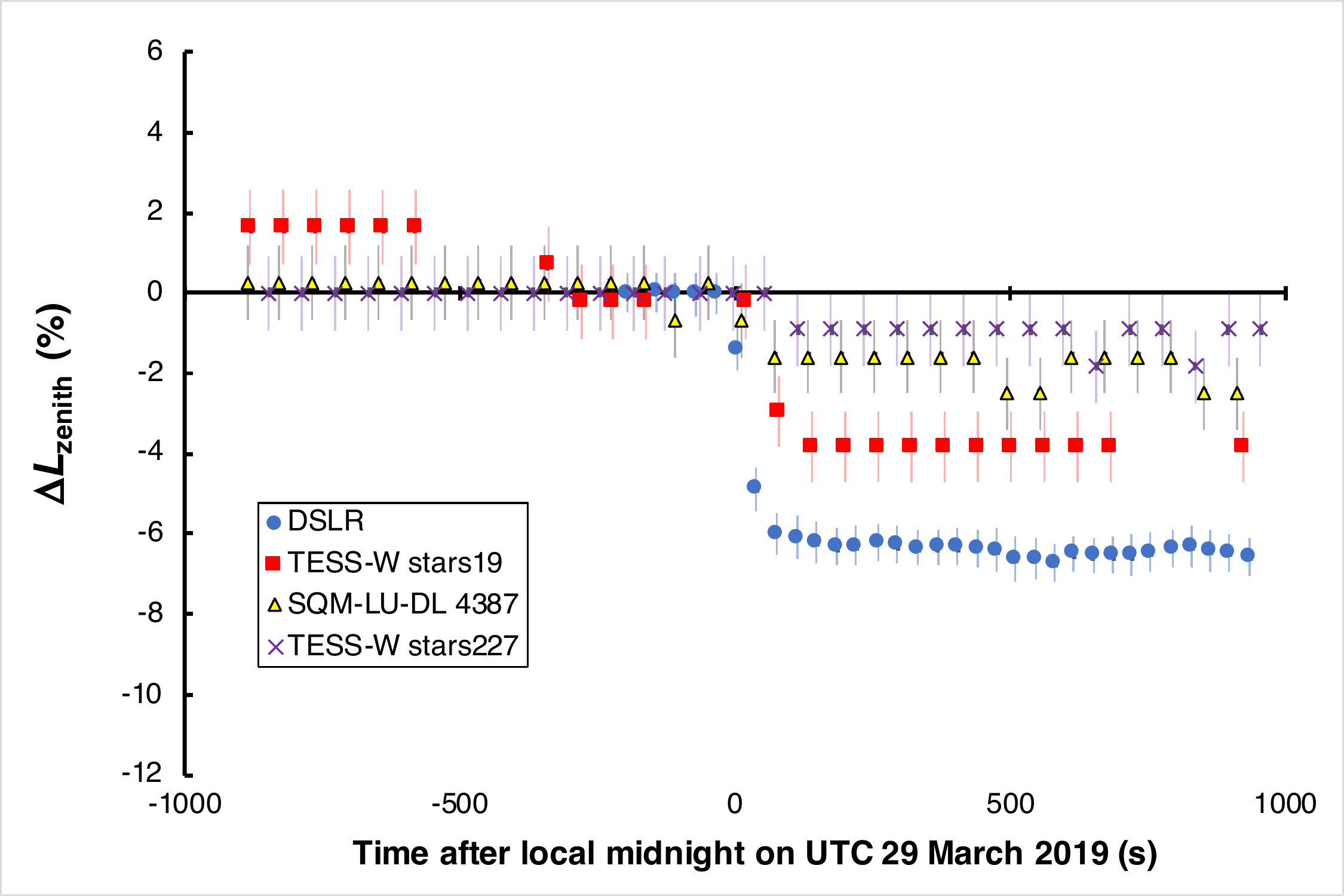}
\caption{Photometry results from the test night of UTC 29 March 2019 during which the Tucson municipal street lighting system was dimmed from 90\% of full power draw before midnight to 30\% after midnight. See main text for details of how the plots were prepared.}
\label{phot-results-29Mar}
\end{figure}

\begin{figure}[htp]
\centering
\includegraphics[width=0.8\textwidth,trim=4 4 4 4,clip]{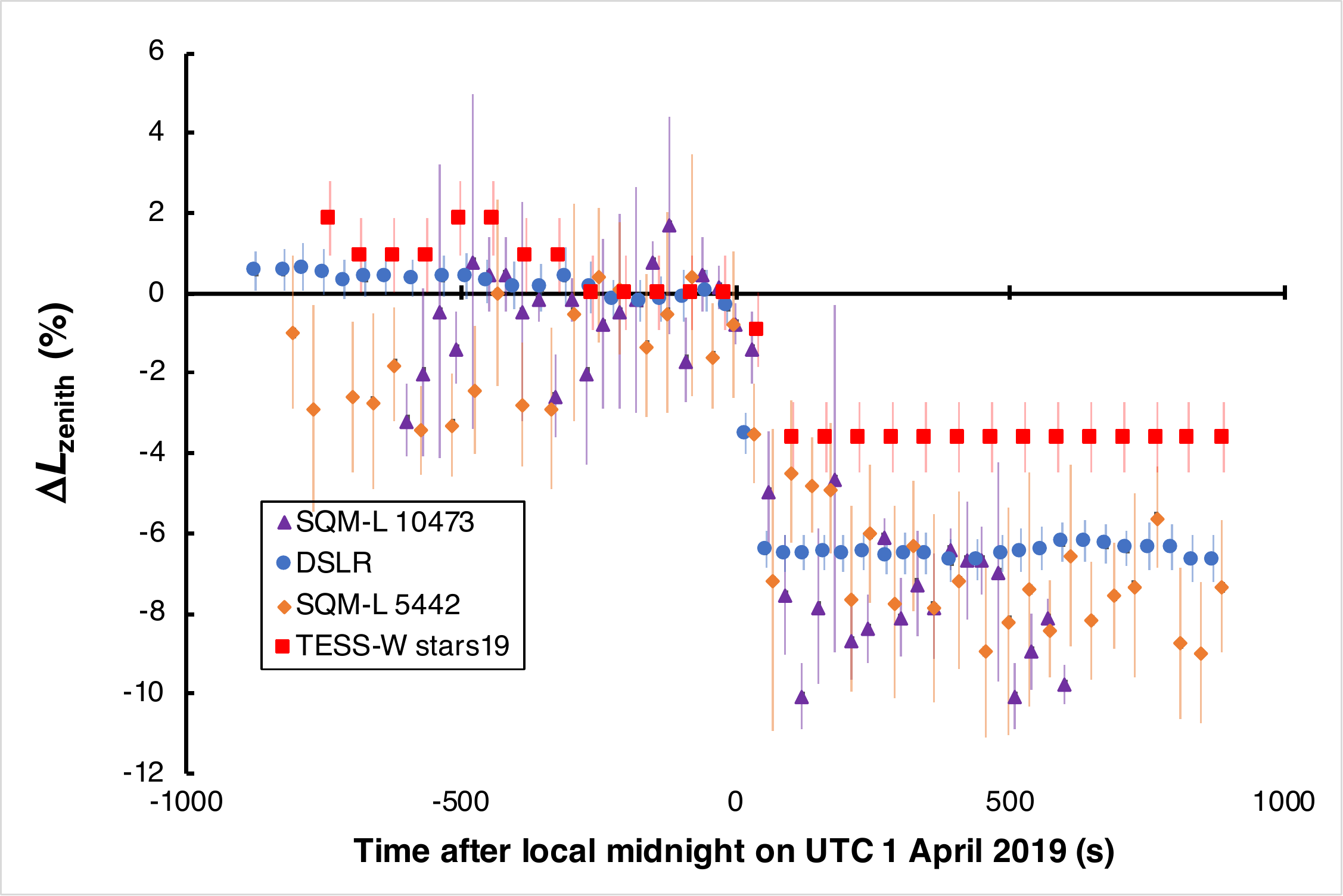}
\caption{Photometry results from the test night of UTC 1 April 2019 during which the Tucson municipal street lighting system was dimmed from 90\% of full power draw before midnight to 30\% after midnight. See main text for details of how the plots were prepared. Note that TESS-W stars227 data are not shown because clouds were present at the installation site on the night of UTC 1 April.}
\label{phot-results-1Apr}
\end{figure}

\begin{figure}[htp]
\centering
\includegraphics[width=0.8\textwidth,trim=4 4 4 4,clip]{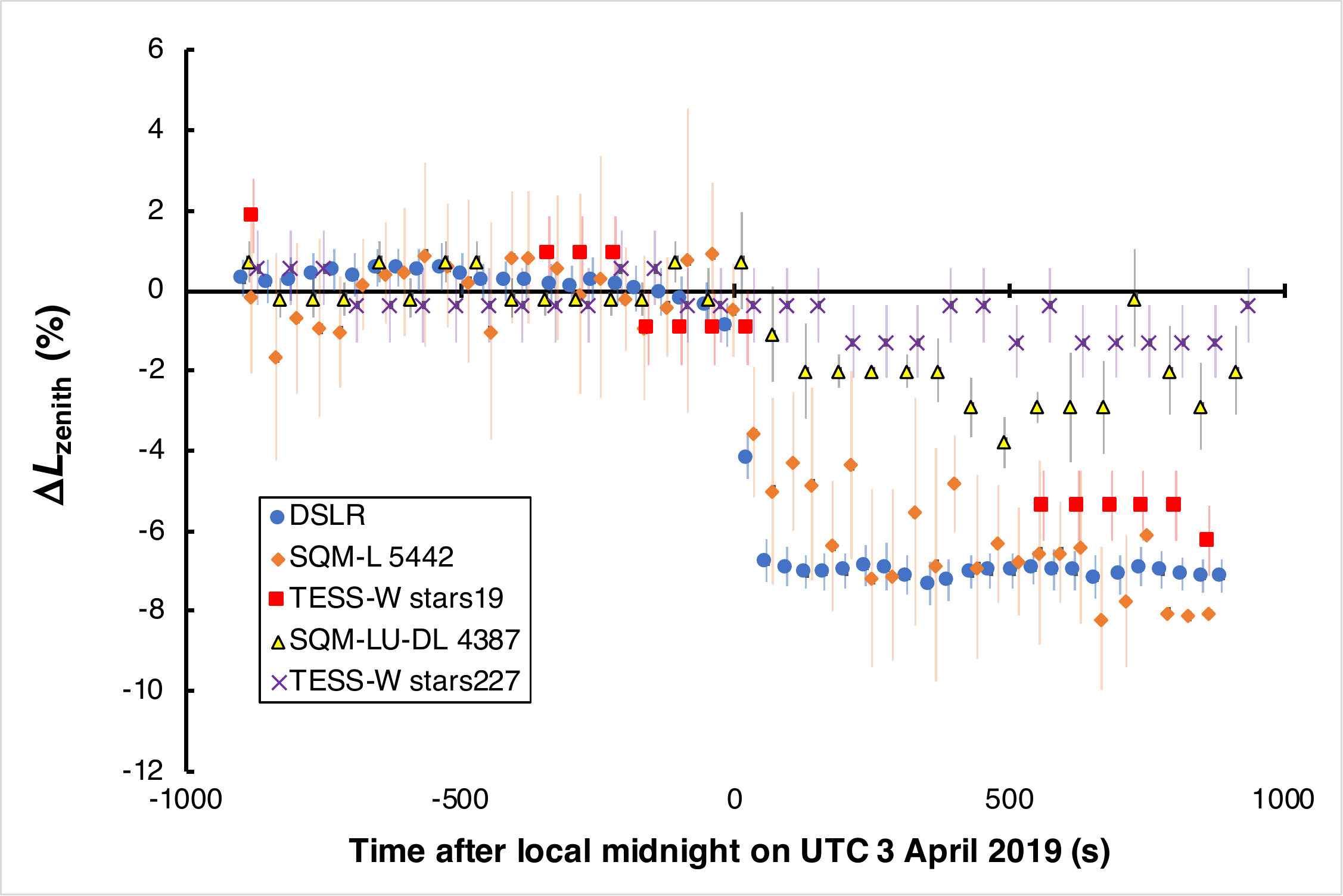}
\caption{Photometry results from the test night of UTC 3 April 2019 during which the Tucson municipal street lighting system was dimmed from 90\% of full power draw before midnight to 30\% after midnight. See main text for details of how the plots were prepared.}
\label{phot-results-3Apr}
\end{figure}

Table~\ref{phot-table} summarizes the measurements from all dimming test nights. Because the analysis presented here focuses on three particular nights among the ten nights for which we obtained photometry, we present in Figures~\ref{phot-results-29Mar}, ~\ref{phot-results-1Apr} and~\ref{phot-results-3Apr} only the time series for the nights of UTC 29 March, 1 April and 3 April 2019, respectively. During these three nights the dimming program sent to the municipal lighting system was to dim from 90\% of the nominal full power draw to 30\% at midnight. These are also the only two nights during the test for which weather conditions were clear and we have a full set of data from all sources. These two nights therefore offer a straightforward interpretation of the data.
\begin{sidewaystable}
\medskip
\resizebox{\linewidth}{!}{%
\begin{tabular}{ ccclccccccc }
 \toprule
\textbf{UTC} & \textbf{Dimming} & \multicolumn{2}{c}{} & \multicolumn{2}{c}{\textbf{Before midnight}} & \multicolumn{2}{c}{\textbf{After midnight}} & \multicolumn{3}{c}{} \\
\cline{5-6}
\cline{7-8}
 \textbf{Date} & \textbf{Program} & \textbf{} & \textbf{} & \textbf{$L_{\textrm{zenith}}$} & \textbf{$L_{\textrm{zenith}}$} & \textbf{$L_{\textrm{zenith}}$} & \textbf{$L_{\textrm{zenith}}$} & \textbf{${\Delta}L_{\textrm{zenith}}$} & \textbf{${\sigma}_{{\Delta}L_{\textrm{zenith}}}$} & \textbf{${\Delta}L_{\textrm{zenith}}$} \\
\textbf{(2019)} & \textbf{(\%)} & \textbf{Location\footnote{Numbers in this column refer to the positions noted in Table~\ref{equipment} and plotted on the map in Figure~\ref{Tucson-data-source-map}.}} & \textbf{Source\footnote{Detailed information about data sources appears in Table~\ref{equipment}.}} & \textbf{(mag arcsec$^{-2}$)} & \textbf{(mcd m$^{-2}$)} & \textbf{(mag arcsec$^{-2}$)} & \textbf{(mcd m$^{-2}$)} & \textbf{(mcd m$^{-2}$)} & \textbf{(mcd m$^{-2}$)} & \textbf{(\%)} \\
\hline
29 Mar	&	30	&	3	&	DSLR		& $18.64\pm0.01$ &	$4.41\pm0.01$ &	$18.71\pm0.01$	& 	$4.14\pm0.01$	&	--0.28	&	0.01	&	--6.3	\\
29 Mar	&	30	&	6	&	stars19		& $20.36\pm0.01$ &	$0.85\pm0.01$	&	$20.41\pm0.01$	&	$0.81\pm0.01$	&	--0.04	&	0.01	&	--4.2	\\
29 Mar	&	30	&	2	&	SQM-L 4387	& $21.50\pm0.01$ &	$0.30\pm0.01$	&	$21.52\pm0.01$	&	$0.29\pm0.01$	&	--0.01	&	0.01	&	--1.6	\\
29 Mar	&	30	&	7	&	stars227		& $21.77\pm0.01$ &	$0.23\pm0.01$	&	$21.78\pm0.01$	&	$0.23\pm0.01$	&	--0.01	&	0.01	&	--0.9	\\
\hline
31 Mar	&	30	&	4	&	SQM-L 10473	& $18.91\pm0.02$ &	$3.20\pm0.07$	&	$18.99\pm0.01$	&	$3.00\pm0.03$	&	--0.20	&	0.08	&	--6.4	\\
31 Mar	&	30	&	6	&	stars19		& $20.32\pm0.01$ &	$0.88\pm0.01$	&	$20.38\pm0.01$	&	$0.83\pm0.01$	&	--0.04	&	0.01	&	--5.1	\\
31 Mar	&	30	&	2	&	SQM-L 4387	& $21.42\pm0.01$ &	$0.32\pm0.01$	&	$21.44\pm0.01$	&	$0.31\pm0.01$	&	--0.01	&	0.01	&	--1.8	\\
31 Mar	&	30	&	7	&	stars227		& $21.67\pm0.01$ & $0.25\pm0.01$ & 	$21.67\pm0.01$ 	&	$0.25\pm0.01$	&	0.00		&	0.01	&	0.0	\\
\hline
01 Apr	&	30	&	3	&	DSLR		& $18.62\pm0.01$ &	$4.51\pm0.01$	&	$18.69\pm0.01$	&	$4.22\pm0.01$	&	--0.29	&	0.01	&	--6.4	\\
01 Apr	&	30	&	3	&	SQM-L 5442	& $18.96\pm0.01$ &	$3.13\pm0.02$	&	$19.02\pm0.01$	&	$2.95\pm0.04$	&	--0.18	&	0.05	&	--5.7	\\
01 Apr	&	30	&	5	&	SQM-L 10473	& $18.86\pm0.01$ &	$3.39\pm0.04$	&	$18.94\pm0.01$	&	$3.14\pm0.06$	&	--0.25	&	0.07	&	--7.4	\\
01 Apr	&	30	&	6	&	stars19		& $20.29\pm0.01$ &	$0.90\pm0.01$	&	$20.34\pm0.01$	&	$0.87\pm0.01$	&	--0.03	&	0.01	&	--3.8	\\
01 Apr	&	30	&	7	&	stars227		& $21.43\pm0.08$ &	$0.30\pm0.01$	&	$21.45\pm0.06$	&	$0.32\pm0.02$	&	+0.02	&	0.02	&	+5.7	\\
\hline
03 Apr	&	30	&	3	&	DSLR		& $18.68\pm0.01$ &	$4.25\pm0.02$	&	$18.76\pm0.01$	&	$3.96\pm0.03$	&	--0.29	&	0.03	&	--6.9	\\
03 Apr	&	30	&	3	&	SQM-L 5442	& $19.10\pm0.01$ &	$2.71\pm0.02$	&	$19.17\pm0.01$	&	$2.56\pm0.04$	&	--0.15	&	0.04	&	--5.7	\\
03 Apr	&	30	&	6	&	stars19		& $20.34\pm0.01$ &	$0.87\pm0.01$	&	$20.39\pm0.01$	&	$0.83\pm0.01$	&	--0.05	&	0.01	&	--5.2	\\
03 Apr	&	30	&	2	&	SQM-L 4387	& $21.35\pm0.01$ &	$0.34\pm0.01$	&	$21.37\pm0.01$	&	$0.34\pm0.01$	&	--0.01	&	0.01	&	--2.0	\\
03 Apr	&	30	&	7	&	stars227		& $21.64\pm0.01$ &	$0.26\pm0.01$	&	$21.65\pm0.01$	&	$0.26\pm0.01$	&	--0.01	&	0.01	&	--1.1	\\
\hline
07 Apr	&	100	&	1	&	SQM-L 4246	& $18.85\pm0.01$ &	$3.44\pm0.04$	&	$18.89\pm0.01$	&	$3.28\pm0.02$	&	--0.16	&	0.05	&	--4.7	\\
07 Apr	&	100	&	3	&	DSLR		& $18.56\pm0.01$ &	$4.72\pm0.01$	&	$18.58\pm0.01$	&	$4.67\pm0.01$	&	--0.06	&	0.01	&	--1.2	\\
07 Apr	&	100	&	3	&	SQM-L 5442	& $19.00\pm0.01$ &	$3.00\pm0.02$	&	$19.01\pm0.01$	&	$2.96\pm0.01$	&	--0.04	&	0.02	&	--1.5	\\
07 Apr	&	100	&	6	&	stars19		& $20.29\pm0.01$ &	$0.91\pm0.01$	&	$20.32\pm0.01$	&	$0.89\pm0.01$	&	--0.02	&	0.01	&	--1.8	\\
07 Apr	&	100	&	2	&	SQM-L 4387	& $21.21\pm0.01$ &	$0.39\pm0.01$	&	$21.23\pm0.01$	&	$0.38\pm0.01$	&	--0.01	&	0.01	&	--1.3	\\
07 Apr	&	100	&	7	&	stars227		& $21.64\pm0.01$ &	$0.31\pm0.01$	&	$21.46\pm0.01$	&	$0.31\pm0.01$	&	0.00		&	0.01	&	0.0	\\
\hline
08 Apr	&	100	&	3	&	DSLR		& $18.73\pm0.01$ &	$4.06\pm0.01$	&	$18.77\pm0.01$	&	$3.93\pm0.01$	&	--0.13	&	0.01	&	--3.1	\\
08 Apr	&	100	&	3	&	SQM-L 5442	& $19.15\pm0.01$ &	$2.59\pm0.01$	&	$19.18\pm0.01$	&	$2.52\pm0.01$	&	--0.08	&	0.02	&	--3.0	\\
08 Apr	&	100	&	6	&	stars19		& $20.29\pm0.01$ &	$0.90\pm0.01$	&	$20.33\pm0.01$	&	$0.87\pm0.01$	&	--0.02	&	0.01	&	--2.7	\\
08 Apr	&	100	&	2	&	SQM-L 4387	& $21.34\pm0.01$ &	$0.34\pm0.01$	&	$21.36\pm0.01$	&	$0.34\pm0.01$	&	--0.01	&	0.01	&	--1.3	\\
08 Apr	&	100	&	7	&	stars227		& $21.58\pm0.01$ &	$0.28\pm0.01$	&	$21.59\pm0.01$	&	$0.27\pm0.01$	&	--0.01	&	0.01	&	--0.6	\\
\bottomrule
 \end{tabular}}
\caption{Summary of photometry results for all test nights with quality ratings of GOOD as indicated in Table~\ref{test-2019-quality}. For each dimming configuration (column 2) the location and data source (columns 3--4) are listed, rows are listed by date according to increasing distance from the Tucson city center. The columns that follow are the mean zenith brightness ($L_{\textrm{zenith}}$) and 1$\sigma$ scatter of measurements before midnight (columns 5--6) and after midnight (columns 7--8), the measured change in zenith brightness across midnight (${\Delta}L_{\textrm{zenith}}$; column 9), rms uncertainty on ${\Delta}L_{\textrm{zenith}}$ (${\sigma}_{{\Delta}L_{\textrm{zenith}}}$; column 10), and ${\Delta}L_{\textrm{zenith}}$ expressed as a percent change relative to $L_{\textrm{zenith}}$ before midnight. To facilitate immediate comparison with values of the night sky brightness in the literature, values of $L_{\textrm{zenith}}$ are given in both magnitudes per square arcsecond and candelas per square meter.}
\label{phot-table}
\end{sidewaystable}


The results shown in the figures involve measurements spanning several decades of zenith brightnesses, so we have prepared the plots in a particular way. First, a constant offset was applied to each time series in which the constant subtracted from all points in the series is the average of values in the final 300 s before midnight. This makes the decrease in zenith brightness from the various sources in differing locations clearer. And second, values on each ordinate are expressed in as a percent change relative to that average brightness in the last five minutes before midnight. 

Brightness calculations were performed in SI luminance units (cd m$^{-2}$). Where data were reported in native logarithmic units of magnitudes per square arcsecond (mag arcsec$^{-2}$), we converted to linear SI units according to the semi-empirical formula
\begin{equation}
L (\textrm{cd m$^{-2}$}) = L_0 \times 10^{-0.4 B_{V}},
\end{equation}
where $B_{V}$ is the sky brightness in units of Johnson-Cousins $V$ magnitudes per square arcsecond~\citep{Bessell1990} and $L_0$ is a constant that depends on the spectral power distribution of the light source. Using the method of Bar\'{a} et al.,~\citep{Bara2020} we find $L_0 = 1.1838\times10^{5}$ cd m$^{-2}$ when referred to the absolute (``AB'') magnitude system.~\citep{Oke1974} We note, however, that the derivation of the scaling factor assumed as input a surface brightness on the Johnson-Cousins $V$ magnitude system, and that the bandpasses of the devices used to estimate the brightness of the night sky are only approximately comparable to $V$.

Uncertainties on the plotted points are estimated from the 1$\sigma$ scatter in each time series in the last 300 s before midnight for all points before midnight, and in the period of 180 to 480 s after midnight for all points after midnight. This accounts for the observed delay in all of the municipal street lights reaching the post-midnight dimming configuration, given that in some instances zenith brightnesses after midnight did not stabilize for up to three minutes after the commanded dimming. In instances where the 1$\sigma$ scatter of points in the time series was less than the 0.01 mag arcsec$^{-2}$ resolution of the photometers we used, the uncertainty was arbitrarily set to $\pm$0.01 mag arcsec$^{-2}$, or about $\pm$0.9\% in luminance. Note that in some cases this range is smaller than the size of the plotted points themselves, and error bars are therefore not always evident in the figures.

The signal attributable to the 30\% dimming program is evident in Figures~\ref{phot-results-29Mar}, ~\ref{phot-results-1Apr} and~\ref{phot-results-3Apr}, along with any changes resulting from uncoordinated dimming of private light sources at the same time. With the exception of the measurement site most distant from the city center (TESS-W unit stars227 at location `7' in Figure~\ref{Tucson-data-source-map} and Table~\ref{equipment}), the zenith brightness drops rapidly after midnight at all sites. However, the drop is not instantaneous; rather, the change is complete after two consecutive measurements in the time series, amounting to 80-120 seconds depending on the data source. Because we believe that the street lighting dimming program was a comparatively strong influence on the zenith brightness change at midnight on these nights, we speculate that the finite time to reach a new zenith brightness just after midnight results from random differences in timekeeping among the luminaires in the street lighting system. The Tucson system receives and transmits information to a central location only at certain times of day, so the execution of the dimming program is carried out according to local timing circuits in each luminaires. If, on the other hand, the execution was centrally timed, we would expect the resulting sky brightness change at midnight to take only one measurement cycle to complete. 

\begin{figure}[htp]
\centering
\includegraphics[width=0.8\textwidth,trim=4 4 4 4,clip]{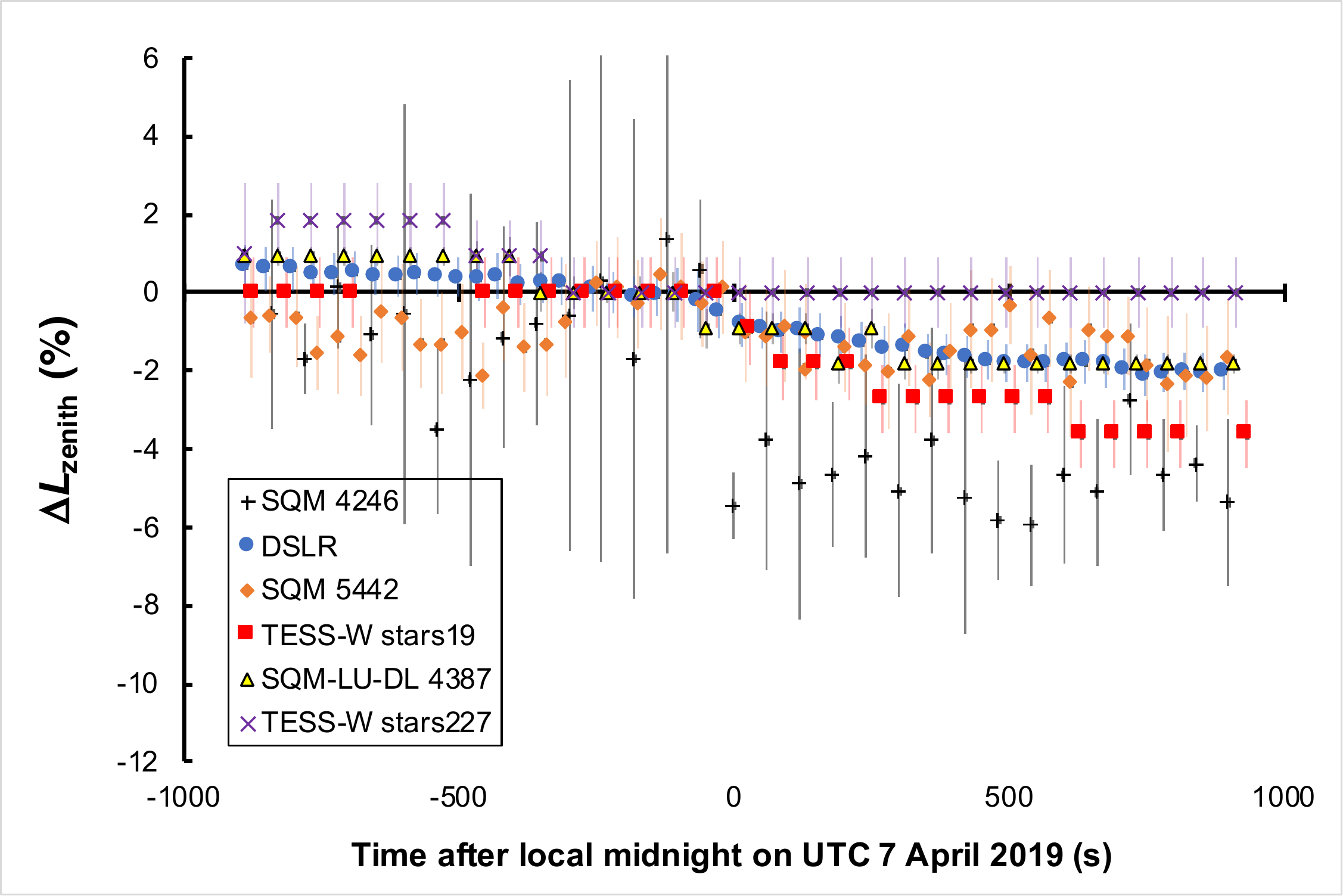}
\includegraphics[width=0.8\textwidth,trim=4 4 4 4,clip]{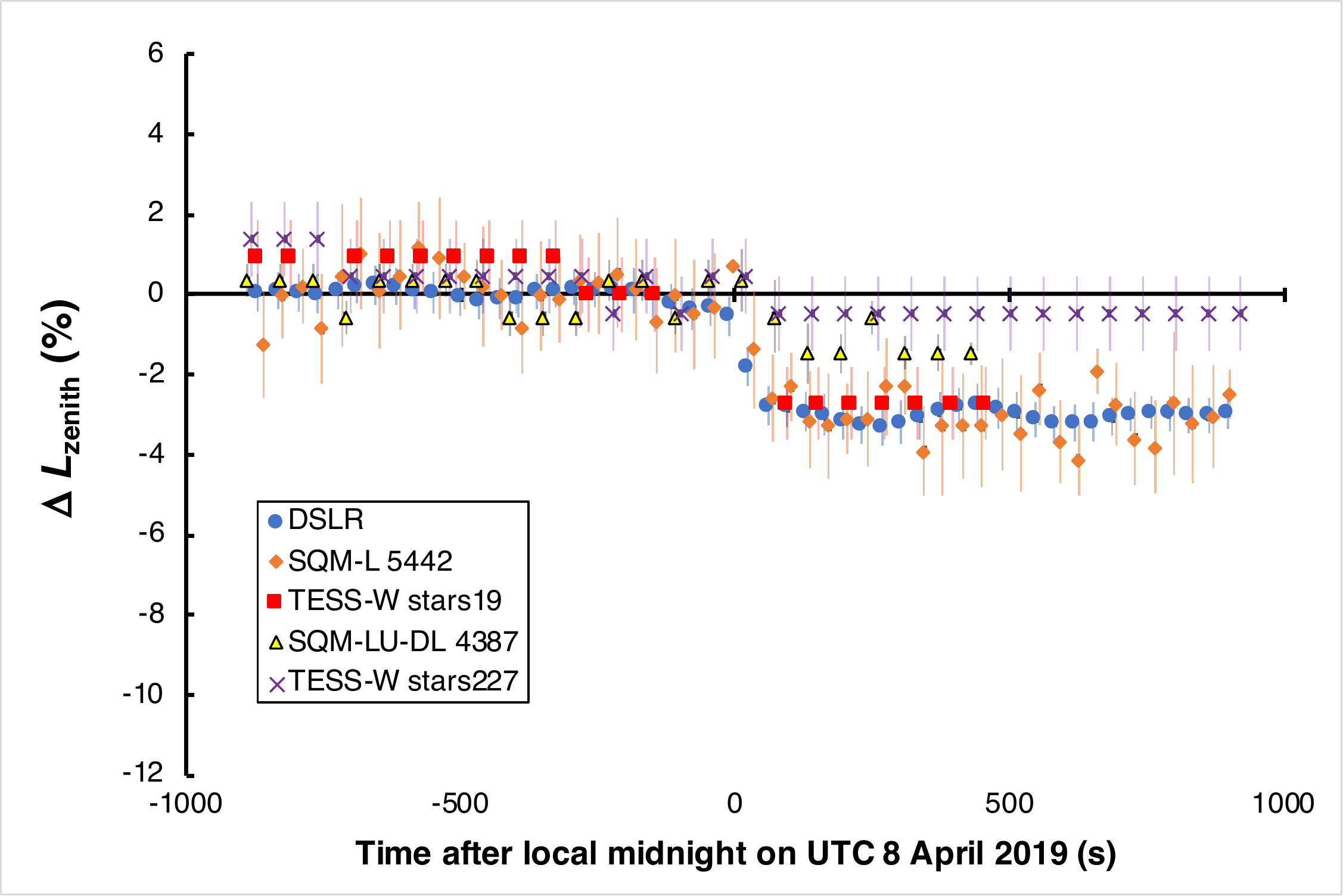}
\caption{Photometry results from the test nights of UTC 7 April and 8 April 2019 during which the power draw of the Tucson municipal street lighting system was increased from 90\% of full power draw before midnight to 100\% after midnight. The plots were prepared in the same manner as those in Figures~\ref{phot-results-29Mar}-\ref{phot-results-3Apr} and are shown on the same vertical scale for purposes of direct comparison.}
\label{phot-results-100}
\end{figure}

We also observed an overall diminution in the brightness of the zenith on test nights when the output of the municipal lighting system was increased from 90\% of the full power draw before midnight to 100\% after midnight. These photometry results are presented in Figure~\ref{phot-results-100}, and were prepared identically to the previous figures. 

Although we expected a gradual change in the sky brightness across midnight, consistent with the trend observed by other authors in time series obtained in cities, we see a more marked diminution.  We interpret these profiles to indicate that Tucson lighting other than known street lighting also decreases at local midnight in an uncoordinated way. Anecdotal observations of privately owned lighting in Tucson suggest that owners of many private properties tend to extinguish certain types of lighting, such as parking lot lighting and illuminated signs, at midnight. The profiles from the two nights shown in Figure~\ref{phot-results-100} therefore represent a drop in sky brightness at midnight due to the routine dimming of private lighting around midnight that us partially `filled in' by the small increase in the output of known street lights from 90\% to 100\% of full power. The shape of the profile differs on the two nights as well: a more gradual change is seen on the night of UTC 7 April, whereas the change on the night of UTC 8 April more closely resembles that seen on the 30\% test nights in Figures~\ref{phot-results-29Mar}-\ref{phot-results-3Apr}. The difference may well result from the fact that night of UTC 7 April was a Saturday, while the following night was a Sunday, given expectations of different patterns of human activities on week nights versus weekend nights.

Comparing Figures~\ref{phot-results-29Mar}-\ref{phot-results-3Apr} and \ref{phot-results-100}, it is clear that the change in sky brightness on nights when the street lighting system was slightly brightened to 100\% of the lights' full power draw is very small. However, using the photometry results for the 30\% power and 100\% power nights, we determined from data collected at two sites within the city that the contribution to the zenith luminance attributable to sources other than known street lights decreases by about 2.5\% at midnight. A derivation of the formulae used to make this calculation is given in the Appendix.

\begin{figure}[htp]
\centering
\includegraphics[width=\textwidth,trim=4 4 4 4,clip]{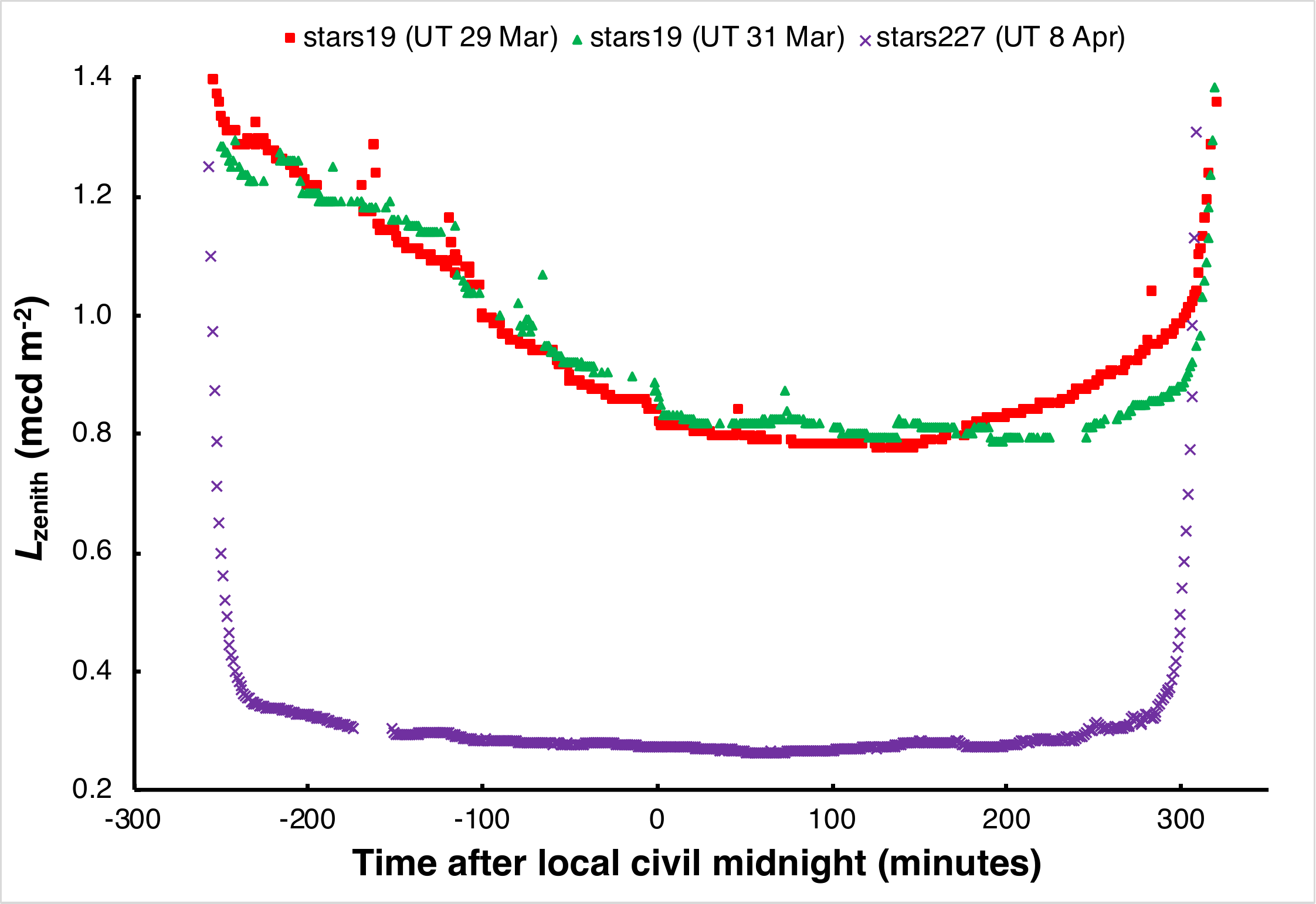}
\caption{Full-night time series data from the TESS-W units used in this study on three clear nights during the Tucson dimming tests. Symbol shapes colors indicate the UT dates of observation: red squares (stars19, UT 29 March), green triangles (stars19, UT 31 March) and purple crosses (stars227, UT 8 April).Note that the divergence between the stars19 traces in the last few hours of the nights plotted is due to the different lunar phase on the two nights and the resulting interference by moonlight.}
\label{tess-full-nights}
\end{figure}
To provide some context for the measurements reported here, we compared data from the two TESS-W units used in this study (stars19 and stars227, representing locations in the city and far from it, respectively) given that they function autonomously and obtain data all night long. For the range of dates during the Tucson dimming tests, we selected the nights with the best weather throughout the night (UT 29 and 31 March for stars19; UT 8 April for stars227) and plotted them together in Figure~\ref{tess-full-nights} after converting from the instruments’ native luminance units (mag arcsec$^{-2}$) to SI units (mcd m$^{-2}$). The abrupt change in zenith brightness at local civil midnight is evident in the stars19 traces, whereas stars227 shows no comparable change at that time. 

\subsection{Brightness attenuation as a function of radial distance}\label{subsec:distance}
\begin{figure}[htp]
\centering
\includegraphics[width=\textwidth,trim=4 4 4 4,clip]{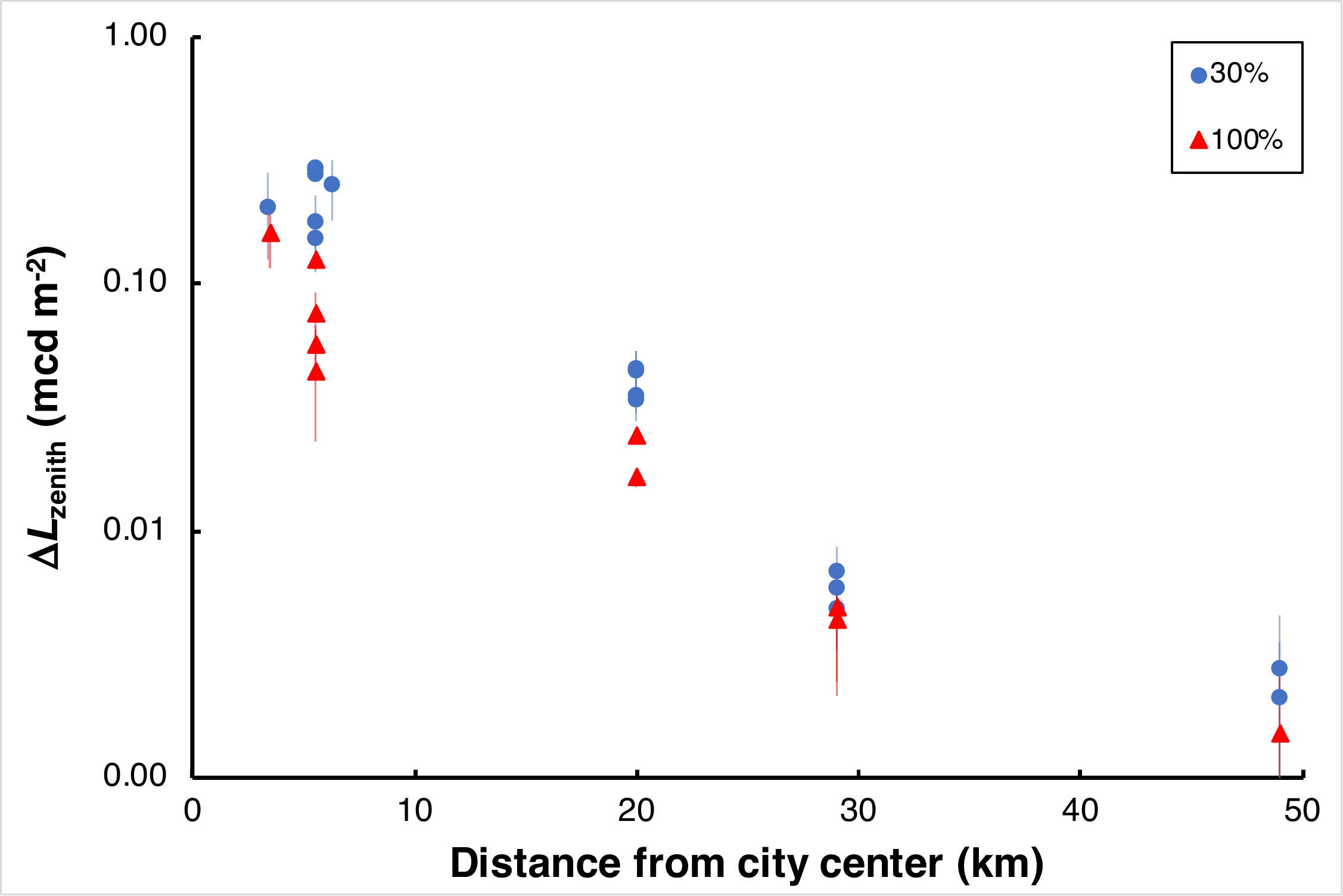}
\caption{Photometry results from the test nights at different post-midnight street lighting system power draw percentages as indicated in the legend at upper right. ${\Delta}L_{\textrm{zenith}}$ is the magnitude of the change in the mean value of zenith brightness at each location across midnight; the sign is dropped to facilitate plotting the ordinate on a log$_{10}$ scale. In this way, the large range of values is more apparent, but the reader should be aware that this approach is different than as shown on the vertical axes of Figures~\ref{phot-results-29Mar}-\ref{phot-results-3Apr} and~\ref{phot-results-100}. Note also that the data plotted here consist only of nights with quality ratings of GOOD as indicated in Table~\ref{test-2019-quality}.}
\label{L-vs-D}
\end{figure}

Our skyglow measurements contain more than just direct evidence of the special street lighting dimming configurations. We find that the magnitude of the change in zenith brightness at midnight is clearly dependent on the radial distance of the measuring station from the city center, as is evident in Figure~\ref{L-vs-D}. The observed zenith luminance changes at midnight, as a function of distance from the city center, are consistent with those reported by Kinzey \emph{et al.}~\citep{Kinzey2017} for the fiducial `near observer' and `distant observer' positions in their models.

The figure suggests that the zenith luminance signal from changes to lighting in the city from fully shielded sources with spectral power distributions like those of the Tucson 3000K white LED street lights is strongly attenuated as a function of distance and asymptotes to a constant value at scales comparable to the linear extent of the city itself ($\sim$20 km). For the most distant measurement station, any change in the zenith brightness across midnight is well within the measurement uncertainty. We take this to be an effective change of zero; the brightness of the zenith at Oracle State Park is therefore not influenced by light from Tucson. We note, however, that this is also the result of topography: the Santa Catalina Mountains (elevation $\sim$2800 m above sea level) are situated directly between the park and Tucson and serve to partially screen out skyglow attributable to city lights.

The curves in in Figure~\ref{L-vs-D} can be fit reasonably with an exponential function whose decay constant is $\sim0.1$ km$^{-1}$, which implies that for this kind of lighting scenario, the \emph{e}-folding rate at which the zenith brightness influence of changes to the street lighting configuration in Tucson declines is about 10 km.

\subsection{Nightly variation in baseline night sky brightness}
It is evident in Table~\ref{phot-table} that the zenith brightness at each of our measurement locations in the several minutes before midnight on each of the test nights varies considerably. The variation from night to night ranges from 8\% near the city center up to $\sim$35\% at the most distant location, Oracle State Park (labeled ``7'' in Figure~\ref{Tucson-streetlight-map}). Although our analysis depended solely on relative changes in sky brightness, as presented in Figures~\ref{phot-results-29Mar}-\ref{phot-results-3Apr} and~\ref{phot-results-100}, we sought to understand why the baseline night sky brightness each night varied by the observed amounts.

As suggested in Section~\ref{subsec:distance}, the influence of anthropogenic skyglow on the zenith night sky brightness declines steadily with increasing distance from the urban source. At the position of Oracle State Park, there is very little artificial light at the zenith and the sky brightness is mainly determined by natural sources of light in the night sky and highly local contributions from nearby sources of artificial light on the ground. We suspected that variations in the natural background account for the observed behavior; indeed, Grauer \emph{et al.}~\citep{Grauer2019} recently showed that temporal trends in night sky brightness at observing stations separated by thousands of kilometers are contemporaneous with solar and geomagnetic activity patterns even during deep solar minima. 

\begin{figure}[htp]
\centering
\includegraphics[width=\textwidth,trim=4 4 4 4,clip]{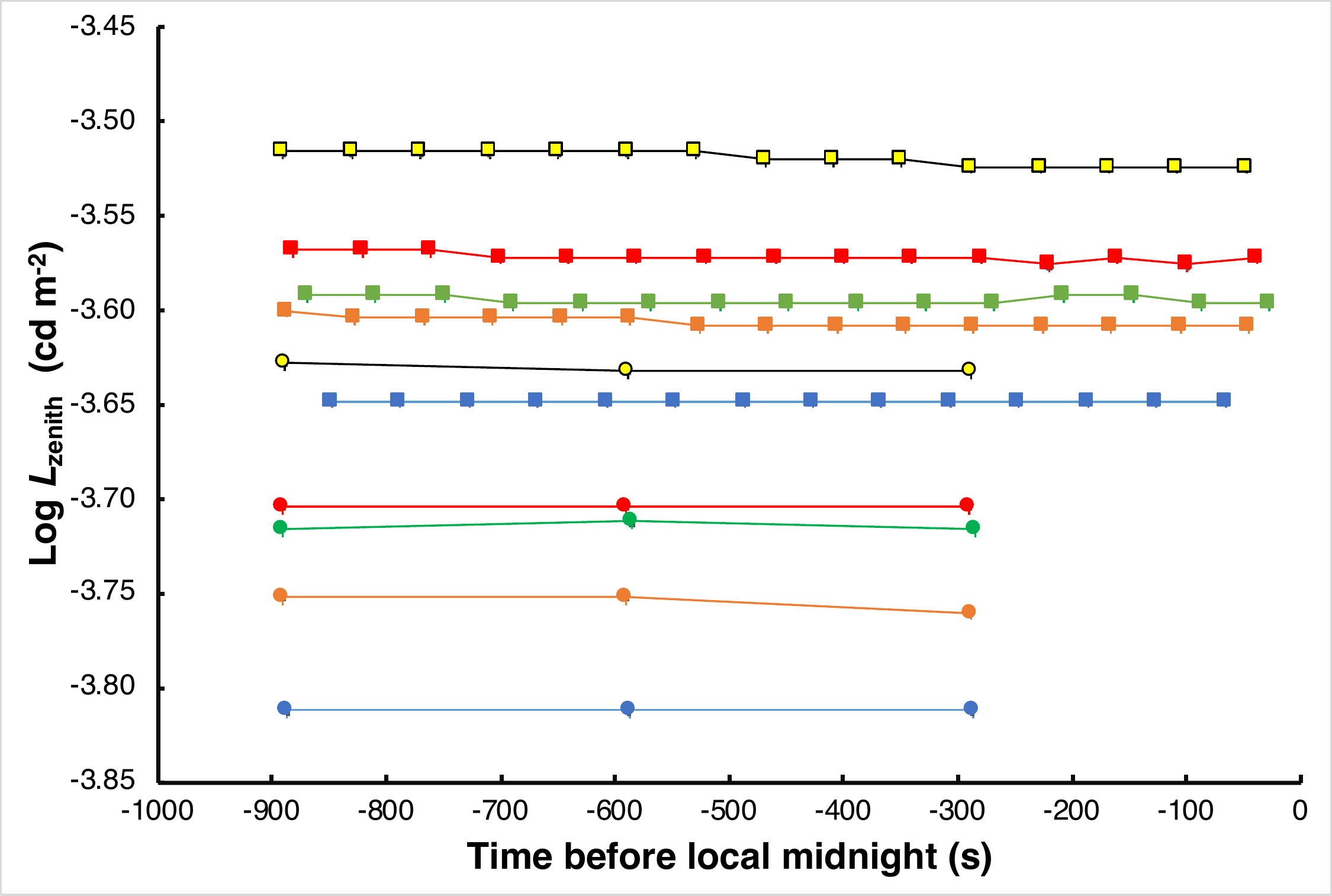}
\includegraphics[width=\textwidth,trim=4 4 4 4,clip]{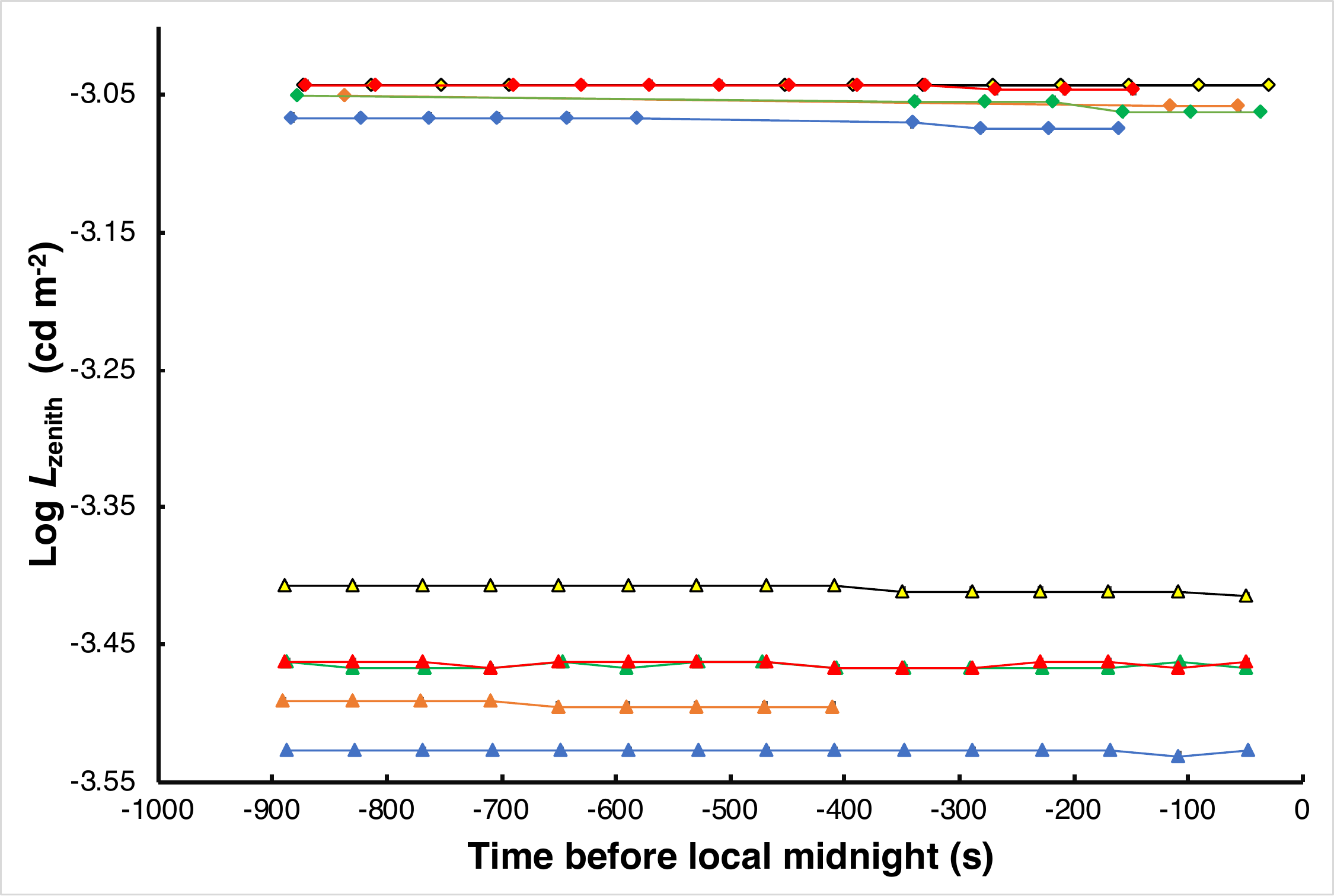}
\caption{Zenith brightness values in the fifteen minutes before midnight (0700 UTC) as seen from four locations on the Tucson dimming test nights. Upper panel: Cosmic Campground (circles) and Oracle State Park (site ``7,'' squares); lower panel: Mount Lemmon (site ``2,'' triangles) and east Tucson (site ``6,'' diamonds). Different UTC dates of observation are denoted by color: 29 March (blue), 31 March (orange), 3 April (green), 7 April (yellow) and 8 April (red). Data points have been joined by connecting lines to guide the eye in discerning between the nights. Although error bars are plotted for all measurements, in each case they are approximately the same size as the symbols themselves.}
\label{nightly-offsets}
\end{figure}
To search for such an effect, we compared contemporaneous TESS-W measurements made at Oracle State Park with SQM-LU-DL data obtained at the Cosmic Campground (33.479551, --108.922529; 1634 m), a protected dark-sky site on U.S.~Forest Service land 235 kilometers to the east-northeast of Oracle State Park. Photometry results for the two sites are shown in the upper panel of Figure~\ref{nightly-offsets}; the data are restricted to the last fifteen minutes before midnight (0700 UTC) on each night in order to eliminate any influence from the Tucson dimming tests on the zenith brightness as seen from Oracle State Park. The zenith brightnesses at the two locations track each other from night to night proportionately: the night of UTC 29 March was the least bright at both locations, while the night of 7 April was the brightest. 

For purposes of comparison, the lower panel of Figure~\ref{nightly-offsets} shows photometry results for two sites closer to Tucson on the same nights. The same general pattern noted in the upper panel is seen here. As the distance to Tucson decreases, the nightly variation in the baseline sky brightness level before midnight decreases in magnitude. We interpret this to result from the contribution from anthropogenic skyglow over the city, which is much less variable from one night to the next than natural sources of light such as airglow. This is consistent with the observation related in the previous section showing the dependence of zenith brightness on distance from the city. Well inside the city, the observed nightly variation in the baseline sky brightness level before midnight does not follow the same pattern. We speculate that this may be due to sensitivity to highly local, variable sources of light on the ground, or to similarly local variations in the atmospheric optical depth.

%
%
\section{SkyGlow Simulator Predictions}\label{sec:simulations}

To model the influence of the municipal street lighting system in the different dimming configurations, we carried out model runs with SkyGlow Simulator (version v.5c)\footnote{The software is publicly available on \url{http://skyglow.sav.sk/#simulator}.} based on theory developed by Kocifaj.~\citep{Kocifaj2007} Given the assumed linear relationship between the number of lumens emitted by the municipal lighting system and the brightness of the night sky at the zenith, and using the known experimental decrease in the zenith brightness each night during the dimming experiment, SkyGlow Simulator results can be used to determine the influence of the known street lights on the zenith brightness. From this we can infer the fraction of light measured at the zenith that is specifically attributable to the municipal street lighting system.

It is important to note that there is no simple correspondence between the fraction of skyglow caused by known street lights and the zenith brightness measured inside the city. Our previous work~\citep{Barentine2018} showed that 43.6\% of known street lights in the Tucson metro area were not replaced by LEDs during the 2016--17 City of Tucson municipal lighting conversion. These lights are owned and operated by municipalities other than the Tucson, or by private entities such as the local electric utility company.  As a result, dimming the new City of Tucson LED street lights from the nominal before-midnight condition of 90\% of the full power draw to 30\% will not yield a 60\% decrease in the zenith brightness. That there are other sources besides street lights makes the expected change in the zenith brightness smaller. Furthermore, because the influence of Rayleigh scattering is a strong inverse function of the wavelength of light, the relationship between zenith brightness and the number of installed lumens in the form of street lighting depends on the spectral power distribution of the light sources. As a result, the dependence of the zenith brightness on the contribution of sources other than street lighting is slightly non-linear. 

\subsection{Model lighting scenarios and inputs}

For this analysis we selected two locations within the municipal boundaries of the City of Tucson: Reid Park (32.21239, -110.92515), and Tanque Verde (32.22388, -110.76741), a suburban census-designated place in northeast Tucson. We restricted the analysis to those two places because Reid Park is near the city center, while Tanque Verde served as something of a control at large radial distance from the city center. 

To find the unknown fraction of lumens indicated by the test conditions, we first computed several models for reference configurations of the street lighting system. For the 100\% case, all city light emissions are accounted for by known street lights only, while for the 50\% case, known street lights account for exactly half of the city emissions. The 0\% case, representing conditions in which the known street light contribution is completely neglected, is implied. Given that the degree of apparent non-linearity is very small, including only these three points (100\%, 50\% and 0\%) resulted in an adequately constrained fit. The 50\% and 100\% configurations were then used to tune the models.

For both Reid Park and Tanque Verde, we computed models for four scenarios:
\begin{enumerate}
\item No additional light sources are present (100\% of skyglow is caused by known street lamps only)
	\begin{enumerate}
		\item Before midnight
		\item After midnight
	\end{enumerate}
\item The same amount of other light sources was added (skyglow is caused 50\% by known street lamps and 50\% by other sources)
\begin{enumerate}
		\item Before midnight
		\item After midnight
	\end{enumerate}
\end{enumerate}

From these three values, 0\%, 50\% and 100\%, interpolation gives the actual fraction matching the change in zenith brightness measured across midnight on the dimming test nights. The third run of each model took into account this calibration, treating the known light contribution to zenith brightness as a free parameter that was varied until the resulting predicted brightness change matched our observations.

To make the simulations as realistic as possible, we used the City of Tucson municipal lighting inventory data to create an effective spectral power distribution of the lighting system, which is a linear combination of the spectrum of 3000 kelvin white LED light and the city spectrum before the LED conversion, consisting mainly of HPS; see \citep{Barentine2018} for details. We assumed that the spectrum of other parts of the city stayed unchanged during the dimming experiment, i.e., that the diversity of lamp types among private lighting sources was the same in 2017 as in 2019, but that there was more private lighting in 2019 proportionate to population growth in the intervening two years.

\subsubsection{Atmospheric parameters\label{aerosol-params}}

We obtained information about the atmospheric optical transmission conditions during the test nights by querying a publicly available database of hourly records from  Pima County Department of Environmental Quality air quality monitoring stations,\footnote{\url{http://envista.pima.gov/}} the U.S. National Aeronautics and Space Administration AErosol RObotic NETwork (AERONET) station in Tucson,\footnote{\url{https://aeronet.gsfc.nasa.gov/}} and weather observations at Tucson International Airport.\footnote{\url{https://forecast.weather.gov/data/obhistory/KTUS.html}} From these sources we derived four of the six main parameters describing the atmospheric aerosol properties on the test nights. Among these are the aerosol optical depth at 500 nm (AOD$_{500}=0.06$) and the \r{A}ngstrom exponent (${\nu}=1.0$), which are related according to AOD $\sim$ ${\lambda}^{-\nu}$. Further, we found the asymmetry parameter (ASY) and single-scattering albedo (SSA) at several fiducial wavelengths, which are given in Table~\ref{aerosols}.

\begin{table}
\centering
\begin{tabular}{ ccc }
 \toprule
\textbf{$\lambda$ (nm)} & \textbf{ASY} & \textbf{SSA} \\
\hline
300 & 0.710 & 0.935 \\
400 & 0.680 & 0.935 \\
680 & 0.640 & 0.935 \\
850 & 0.635 & 0.935 \\
1050 & 0.635 & 0.935 \\
\bottomrule
 \end{tabular}
\caption{Asymmetry parameter (ASY) and single-scattering albedo (SSA) for the atmosphere over Tucson during the test nights in the experiment obtained from an analysis of air quality and weather data.}
\label{aerosols}
\end{table}

For the remaining two parameters we took standard values: 8.0 km for the scale height of molecular atmosphere and 0.65 km$^{-1}$ for the vertical gradient of aerosol concentration. Lastly, for the surface albedo, we used the same average value (0.15) for Tucson as in \citep{Barentine2018}, which was computed from VIIRS-DNB data obtained in 2017.

We assumed that values of ASY, SSA, and AOD did not change appreciably from one night to the next during the experiment. ASY and SSA are expected to vary locally according to changing weather conditions. We checked the numerical values to look for any large variations or trends that would influence the models. For the nights of UTC 31 March, and 2 and 3 April, the value of ASY was unchanged, and SSA changed only by a small amount, from 0.95 to 1.00, between UTC 2 April and 3 April. The same is true for AOD, which was practically unchanged between UTC 31 March and 4 April. 

Our models evaluate only relative changes across each night at local midnight, and these relative changes are only slightly sensitive to ASY, SSA and AOD. If light scattering in the atmosphere is higher on one night compared to the next, it would be evident as a decrement in the magnitude of the observed sky brightness change at midnight. We therefore conclude that (1) there is no utility in using different values of ASY, SSA and AOD calculated for each night of the experiment; and (2) any changes in the values of these parameters in the $\sim$30 minutes around midnight of each night during which we measured sky brightnesses have significantly smaller effects on the photometry results than do the resolutions of the measuring instruments and the random influences of other sources, such as contributions to skyglow from sources of light other than known street lights. 

\subsubsection{Lighting system configuration}\label{subsec:config}

During the experiment, the City of Tucson provided daily summaries of data returned from its network-enabled street lighting system indicating the actual performance of each individual luminaire on the previous night. This allowed us to determine the  power draw of the system when operating at 100\% of nominal full power ($\sim$1.62 MW). It also showed the fraction of non-communicating luminaires; on average, 6.7\% of lamps, totaling 109 kW, returned power draw values of either 0\% or 100\% of full power during the test nights. We interpreted these figures to mean that either these luminaires did not respond to the commanded dimming sequence or simply failed to operate, respectively. 

To allow for their presence in the simulations, we assumed that their real power draw was 50\% of nominal full power, so that they drew $\sim$54 kW. The correction is not critical to the interpretation of the simulation runs because its influence is below the detection limit of our photometers. The correction is therefore half of 6.7\%, or 3.3\%, of the total installed power in the street lighting system. In comparison, our photometric detection limit is 0.01 mag arcsec$^{-2}$, or a luminance difference of about 0.9\%. This amounts to between 14\% and 25\% of the measured decrease in zenith brightness during the tests on the 30\% power nights.

\begin{table}
\centering
\begin{tabular}{*{3}{ccccc}}
 \toprule
& \multicolumn{2}{c}{\textbf{Before midnight}} & \multicolumn{2}{c}{\textbf{After midnight}} \\
\cline{2-3}
\cline{4-5}
 & \textbf{Power} & \textbf{Light } & \textbf{Power} & \textbf{Light} \\
\textbf{UTC Date} & (MW) & (Mlm) & (MW) & (Mlm) & \\
\hline
29 Mar 	& 	1.363	& 161	& 0.753	& 93	 	\\
1 Apr 	& 	1.375 	& 162 	& 0.718 	& 89	 	\\
3 Apr 	& 	1.371 	& 162 	& 0.707 	& 88 		\\
\bottomrule
\end{tabular}       
\caption{The total power draw and light emissions of the street lighting system on the test nights during which the system power draw was reduced from 90\% of nominal full power before midnight to 30\% of nominal full power after midnight.}
\label{lighting-info}
\end{table}
From the nightly reports from the lighting system, we estimated the maximum power draw before and after midnight on the 30\% test nights. Using the detailed inventory of the municipal lighting system, we calculated mean luminous efficacies for the system in different dimming configurations, finding a value of 118 lm W$^{-1}$ when the system is in the nominal pre-midnight configuration drawing 90\% of full power, while for dimming values below 60\% of full power we find a value of 124 lm W$^{-1}$. Using these efficacies, the number of lumens emitted by the lighting system corresponding to the dimming configurations was calculated. The results are reported in Table~\ref{lighting-info}.

For all other known light sources in the city (e.g., non-retrofitted municipally owned lamps in the city center, and private lighting), we took values from the light-clustering analysis in \citep{Barentine2018} based on VIIRS-DNB observations from 2017.

\subsection{Results of model runs}

\begin{table}
\centering
\begin{tabular}{*{3}{lcc}}
 \toprule
 & \textbf{Known street lighting} & \\
\textbf{Location} & \textbf{contribution (\%)} & \textbf{${\Delta}L_{\textrm{zenith} }$ (\%)} \\
\bottomrule
Reid Park & 25$\pm$4 & --5.4$\pm$0.9 \\
Tanque Verde & 27$\pm$4 & --3.6$\pm$0.9 \\
\end{tabular}       
\caption{Results of model runs showing predicted changes in the brightness of the night sky at the zenith (${\Delta}L_{\textrm{zenith}}$) as a fraction of the contribution of known municipal street lighting to the total city light emission on nights during which the known lighting was dimmed from 90\% of the nominal full power draw to 30\% during overnight tests on 1 April and 3 April 2019. The known lamp contribution in the model was adjusted until the experimentally determined value of the zenith brightness change was obtained.}
\label{model-runs}
\end{table}

Outcomes from the simulations are summarized in Table~\ref{model-runs}. For the nights on which known street lights were dimmed to 30\% of their full power draw after midnight, we calculated the expected decrease in zenith brightness (${\Delta}L_{\textrm{zenith}}$) as a function of the fraction of the city's total light emission attributable specifically to the known street lights. The contribution of known lights in the second table entry for each site should be a simple linear extrapolation, but the known non-linearity of the relationship between known street light fraction and expected zenith brightness discussed previously introduces a small quadratic term in the fit. The extra term in the improved fit raises the relative fractions slightly, from 20\% to 25\% at Reid Park, and from 24\% to 27\% at Tanque Verde. The difference resulting from the added quadratic term is small compared to skyglow measurement uncertainty ($\pm$0.9\%). Applying this uncertainty to the interpolation procedure described above yields an error in the estimate of the contribution to skyglow of known street lights of about $\pm$4\%.

The range of the resulting known-light fractions results from the fact that conditions were not precisely the same on each test night: the responses of the street lighting system to the commanded dimming program varied according to the figures in Table~\ref{test-percentages}, and the distribution of aerosols in the atmosphere varied from night to night relative to the averages given in Section~\ref{aerosol-params}. The range is therefore a first-order estimate of the uncertainties on the inferred known-light contributions. For the analysis in the next section, we used the arithmetic mean value of (26$\pm$4)\% to indicate the modeled fraction of the total city light emission represented by the known street lights.

%
%
\section{Analysis and Discussion}\label{sec:analysis}

We hypothesized the following in terms of the expected results: 
\begin{enumerate}
\item The relative decrease in the zenith brightness on the 30\% dimming nights should be considerably lower than the $\sim$10\% change observed previously as a consequence of the LED lighting conversion in 2016--17.~\citep{Barentine2018}  The reason for this is that the fraction of lumens emitted by the LED luminaires after conversion is substantially lower than that emitted by the legacy (mostly) HPS lighting system. Reducing the output from a smaller starting number of lumens during the dimming experiment should therefore yield a proportionately smaller change in skyglow over the city than the bulk reduction of light emission during the LED conversion itself. 

\item The amount of skyglow attributable to known municipal roadway luminaires should be considerably less than 50\% of total skyglow. There are two reasons for this expectation. First, again, the number of lumens of light emitted by municipal lighting was significantly reduced during the LED conversion. And second, the amount of privately owned lighting grew at an unknown rate since the municipal LED conversion in some way that probably relates to the overall rate of Tucson population growth, which was about +0.75\% year$^{-1}$ in 2017 and 2018.~\citep{census} We therefore considered any change in private lighting since 2017 to be negligible in the analysis of the data presented here.
\end{enumerate}

The observed changes in the zenith luminance from all locations during the dimming tests were near the detection limits of both the SQM and TESS devices ($\sim$0.01 mag arcsec$^{-2}$; $\sim$0.02 mcd m$^{-2}$), so we decided to average together measurements from all devices and clear test nights in each of the two dimming configurations from two of the locations. On the nights during which lights were dimmed to 30\% of full power, we found for Reid Park a zenith luminance change (${\Delta}L_{\textrm{zenith}}$) after midnight of $-0.24\pm0.06$ mcd m$^{-2}$ ($-5.4$\%), while for Tanque Verde we measured $-0.04\pm0.01$ mcd m$^{-2}$ ($-3.6$\%). Based on the model runs discussed in the previous section, these figures correspond to circumstances in which the known municipal street lighting contributes ($26\pm4$)\% of the total city light emission. 

Given this fraction and our estimate of the total Tucson metropolitan area light emissions attributable to street lighting after the conversion of City of Tucson street lights to LED, we can estimate the total light emission of the metropolitan area from all sources. After the LED conversion, we find that total metropolitan area light emissions attributable specifically to street lighting was 1340 Mlm. This consists of 790 Mlm of street lighting not operated by the City of Tucson, 371 Mlm of mainly HPS street lighting operated by the City of Tucson that did not undergo conversion to LED, and 179 Mlm of dimmable 3000K LED street lights. The implied total light emission of the metropolitan area from all sources is therefore 5150$^{+210}_{-190}$ Mlm. 

Private lighting may account for fewer total emissions than the figures above imply for two reasons. First, our simulations do not take into account the near-horizontally directed emissions from commercial lighting, light escaping from building interiors through windows, automotive lighting, illuminated advertising, etc.~\citep{Dobler2015,Dobler2016} These sources emit substantially into the upper hemisphere, so that a relatively small quantity of light yields significant skyglow.  And second, many of the same sources produce significantly more short-wavelength light than the municipal street lights. Since short-wavelength light scatters more strongly during its transit through the atmosphere than long-wavelength light, again a smaller quantity of light can have a disproportionately large effect on skyglow, especially as seen at large distances from the source. If the spectral power distributions of the non-street lighting sources in Tucson differs significantly from HPS, the percentage quoted above would change.

Additionally, our models do not account for the change in zenith brightness at midnight due solely to sources other than known street lighting, which tends cause overestimation of the contribution to zenith brightness from known street lighting. The decrease in zenith brightness on the two 100\% test nights analyzed here varied from no apparent change on UTC 7 April to a change of $-0.08$ mcd m$^{-2}$ on UTC 8 April. Because the slight increase in light emissions on the 100\% test nights represent an almost trivial change in the output of the fully shielded street lighting system, the observed decrease in zenith brightness across midnight on these nights is a rough guess of the fractional amount by which private lighting emissions change at midnight. Indeed, it is consistent with the $\sim$2.5\% decrease in zenith brightness attributable to private lighting computed independently in the Appendix. The speculative correction of somewhere between $0.00-0.08$ mcd m$^{-2}$ for private lighting, given the related increase in experimental uncertainty, changes our modeled contribution to zenith brightness due to known street lights. In the case of no contribution by private lighting, the estimate remains 30\%. For the observation on UTC 8 April in which private lighting dimmed or switched off at midnight results in a $-0.08$ mcd m$^{-2}$ in the zenith brightness, the corresponding street light contribution is 13\%. Experimental uncertainties as high as 30\% in sky brightness measurements in cd m$^{-2}$ (Table~\ref{phot-table}) simply do not allow us to draw stronger conclusions. 

Lastly, another approach to estimating the street light contribution to zenith brightness is implied by the derivation in the Appendix. Expressions are provided for the fraction of zenith brightness attributable to known street lighting before midnight ($L_{S,b}$) relative to the total measured zenith brightness at a given location ($B$) under the two dimming conditions, 30\% and 100\%. These expressions depend only on the measured change in zenith brightness across midnight (${\Delta}L_{z}$); $B$; the fractional change in known street light output in the tests ($X$); and the relative dimming of private light sources across midnight ($Y$), which we experimentally determined to be $-2.5$\%. As expected, we found that the resulting fraction $L_{S,b}$/$B$ depends on location and is higher near the city center, ($3\pm1$)\% at Reid Park, and lower near the edge of the city, ($2\pm1$)\% at Tanque Verde.

There is clear disagreement between the percentage contribution to zenith brightness due to known street lighting derived from radiative transfer modeling of skyglow (14\%) and direct measurement (2--3\%). The difference between the modeled and experimentally determined values from skyglow may be explained if our radiative transfer model underestimates the total city light emission. The model incorporates little information about private lighting, and does not account for transient sources of light whose emission patterns are near-horizontal, such as interior lighting escaping through uncovered building windows, automotive lighting, and illuminated signs. If this is the case, it suggests that these sources of light contribute more to the brightness of the night sky at the zenith as seen from in and near cities than previously understood. Further work is needed to fully explain the disagreement among the results from these approaches.

\subsection{Limitations of this work}

There are a number of factors in the experimental design and execution that could be improved in future dimming tests. 

\begin{enumerate}
\item Because such tests effectively integrate the contributions of all light sources on the ground over distance scales of kilometers, skyglow measurements cannot completely separate the contributions of public and private lighting. This is as compared to remote sensing data, which in concert with land use information could be used to more readily detect the light specifically attributable to one form of lighting or the other.
\item The small non-linearity in the estimation of the amount of ``unknown'' light sources discussed previously introduces some error into the determination of the street lighting fraction.
\item In this study we assumed that the behavior of private and commercial lighting in Tucson at midnight is the same from night to night, which does not take into account human activity patterns involving the use of outdoor lighting on weeknights versus weekends.~\citep{Dobler2015} 
\item Tucson is not an isolated conurbation representing a single municipal entity, but rather is bordered by several neighboring suburbs and completely encloses one (City of South Tucson). These suburbs provide their own public street lighting, as does surrounding Pima County in certain parts of unincorporated territories adjacent to Tucson. While have no reason to believe any of these lighting systems behaved in any particular way on the test nights reported here, we cannot rule out some unknown influence on our sky brightness measurements.
\item The reliability of the power draws reported back by the Tucson municipal lighting system in relation to their light output was not tested, although Kyba \emph{et al}.~\citep{Kyba2019} concluded from remote sensing data that the influence of any departure between commanded and achieved light output states was small.
\end{enumerate}

%
%
\section{Summary and Conclusions}\label{sec:summary}

We made photometric measurements of the brightness of the night sky over Tucson, Arizona, U.S., during a series of nights in March and April 2019 during which the municipal street lighting system was dimmed to a set of non-standard configurations. The experiment was designed to sense the dimming signal in the skyglow over the city, and the resulting measurements were compared against a radiative transfer model of skyglow to recover the fraction of total city light emissions specifically attributable to street lighting, for which a highly complete inventory is available. 

Our SkyGlow Simulator model can successfully explain the experimentally observed decrease in the brightness of the night sky at the zenith during the municipal street lighting experiment in Tucson. The model of light sources, consisting of 26\% known street lights and 74\% of other sources, can be used in the future to predict the brightness of the night sky in various locations in and around the city and under different atmospheric conditions. Extrapolating the observed zenith brightness change for a given decrease in the power draw of the street lighting system to a scenario in which the entire system were fully extinguished implies that street lights contribute about 14\% of the observed brightness of the zenith as seen from Reid Park and Tanque Verde. The observations reported here further imply that the contribution to zenith brightness represented by sources of light other than known street lighting decreases by about $-2.5$\% at midnight on most nights.

The small observed changes in zenith brightness during the tests are caused by the fact that after the conversion of municipal street lighting from mainly HPS to 3000 kelvin white LED in 2016--17 the dimmable LED luminaires contribute a relatively small fraction of the total city light emissions. Therefore, dimming the LED luminaires from 90\% of their full power draw to 30\% has only a small influence on skyglow over the city. This confirms that the modernization of the lighting system in Tucson was successful from a lighting design perspective; however, it also suggests that cost savings and skyglow reductions attributable specifically to dimming programs are relatively small compared to the potential change involved if total light system emissions are reduced during conversions of municipal street lighting systems from legacy technologies to solid state sources. The changes to the night sky luminance resulting from the changes during the experiment described here are so small as to be imperceptible to the human eye, yet many municipalities may find value in actively dimming street lighting even after substantially reducing the overall light emissions in the transition to LED.

Given the results of this study, we speculate that further reductions in zenith brightness over Tucson can be achieved by:
\begin{enumerate}
\item Replacing commercially owned lighting, consisting of mainly HPS sources, with modern LEDs, along with a concomitant reduction in lumens along the lines of the one-third reduction that accompanied the transition of the municipal lighting system to LED; and
\item Optimizing other light sources, consisting of largely architectural and advertising sources. The prediction of the result of such optimization would require characterization of existing light sources installed in Tucson, which can be done by an advanced analysis of the spectrum of the night sky over the city. 
\end{enumerate}

\section*{Appendix}

The luminance of the night sky is the sum of contributions from all sources of light, both natural and artificial. In the context of a city like Tucson, we assume that anthropogenic sources of light dominate the sky luminance, and that natural sources of light are small enough to neglect in the derivation that follows here.

The change in zenith luminance across midnight during the experiment described in the main text, ${\Delta}L_{z}$, is defined as the difference between the luminance after midnight, $L_a$, and the luminance before midnight, $L_b$. In referring to these conditions, we use the subscripts $a$ (after midnight) and $b$ (before midnight). In this formalism,
\begin{equation}
{\Delta}L_{z} = L_a - L_b.
\end{equation}

To this point, the luminances are the totals from all light sources; we assume that the relationship between the city light output from all sources and skyglow is linear. We assert that the luminance can be separated into two components, one representing known street lighting, $S$, and another representing all lighting \emph{other} than known street lighting, which we refer to as ``private'' lighting, $P$. These subscripts can be combined with $a$ and $b$ to denote the zenith luminance contributions of four specific lighting configurations:

\begin{itemize}
\item[$L_{S,a}$] = luminance contribution from known street lighting after midnight
\item[$L_{S,b}$] = luminance contribution from known street lighting before midnight
\item[$L_{P,a}$] = luminance contribution from ``private'' lighting after midnight
\item[$L_{P,b}$] = luminance contribution from ``private'' lighting before midnight
\end{itemize}
\noindent
${\Delta}L_{z}$ can therefore be written according to the contributions of these four sources.
\begin{subequations}
\begin{align}
{\Delta}L_{z} = (L_{S,a} + L_{P,a}) - (L_{S,b} + L_{P,b})
\label{eqn3a} \\
{\Delta}L_{z} = (L_{S,a} - L_{S,b}) + (L_{P,a} - L_{P,b})
\label{eqn3b}
\end{align}
\end{subequations}

We define two constants that represent the relative dimming of each source across midnight as ratios of luminances:
\begin{subequations}
\begin{align}
X \equiv \frac{L_{S,a}}{L_{S,b}}\\
Y \equiv \frac{L_{P,a}}{L_{P,b}}
\label{eqn4b}
\end{align}
\end{subequations}
where $X$ and $Y$ are always positive. $X$ is known for each night during the experiment; we wish to find $Y$, which is otherwise unknown. ${\Delta}L_{z}$ can be expressed in terms of these two constants:
\begin{subequations}
\begin{align}
{\Delta}L_{z} & = (XL_{S,b} - L_{S,b}) + (YL_{P,b} - L_{P,b})\\
 & = L_{S,b}(X-1) + L_{P,b}(Y-1).
\label{eqn-five}
\end{align}
\end{subequations}

On nights that the known street lights dimmed from 90\% to 30\% of full power at midnight, the fractional change in their output is $\frac{3}{9}$, so $X=0.\bar{3}$. We refer to this condition as $X_{30}$. On nights that the known street lights brightened from 90\% to 100\% of full power at midnight, the fractional change in their output is $\frac{10}{9}$, so $X=1.\bar{1}$. We refer to this condition as $X_{100}$. We measured the total luminance of the zenith before midnight under both conditions, which we call $B$:
\begin{equation}
B \equiv L_{S,b} + L_{P,b}.
\label{total-bef}
\end{equation}

Finally, we assume that the fractional change in the output of ``private'' lighting is constant from one night to the next, so $Y$ gets no subscript. There is cause to believe that $Y$ may depend on the day of the week, as there is generally more human activity to later hours in American cities on Friday and Saturday nights than on other nights of the week. Our analysis avoids relying on data from either Friday or Saturday nights for that reason.

A system of two equations depending only on conditions before midnight therefore describes the two dimming conditions:
\begin{subequations}
\begin{align}
{\Delta}L_{z,30}   & = L_{S,b}(X_{30}-1) + L_{P,b}(Y-1)\\
{\Delta}L_{z,100} & = L_{S,b}(X_{100}-1) + L_{P,b}(Y-1).
\end{align}
\end{subequations}
The system can be cast in terms of $L_{P,b}$ by substituting $B$ from Equation~\ref{total-bef} into Equation~\ref{eqn-five} and rearranging:
\begin{subequations}
\begin{align}
{\Delta}L_{z,30}   & = (B - L_{P,b})(X_{30}-1) + L_{P,b}(Y-1)\\
{\Delta}L_{z,100} & = (B - L_{P,b})(X_{100}-1) + L_{P,b}(Y-1).
\end{align}
\end{subequations}
Expanding the equations and canceling terms gives
\begin{subequations}
\begin{align}
{\Delta}L_{z,30}   & = B(X_{30} - 1) + L_{P,b}(Y - X_{30})\\
{\Delta}L_{z,100}   & = B(X_{100} - 1) + L_{P,b}(Y - X_{100}).
\end{align}
\end{subequations}
These equations can be rearranged to result in an expression for $L_{P,b}$:
\begin{subequations}
\begin{align}
{\Delta}L_{z,100} - B(X_{100} - 1) & = L_{P,b}(Y - X_{100})\\
L_{P,b} & = \frac{{\Delta}L_{z,100} - B(X_{100} - 1)}{(Y - X_{100})}.
\end{align}
\end{subequations}
The result is substituted into the previous expression for ${\Delta}L_{z,30}$:
\begin{equation}
{\Delta}L_{z,30} = B(X_{30} - 1) + \frac{{\Delta}L_{z,100} - B(X_{100} - 1)}{(Y - X_{100})} (Y - X_{30}),\\
\end{equation}
and rearranged to solve for $Y$:
\begin{subequations}
\begin{align}
{\Delta}L_{z,30} - B(X_{30} - 1) & = [{\Delta}L_{z,100} - B(X_{100} - 1)]\left(\frac{Y - X_{30}}{Y - X_{100}}\right)\\
\frac{{\Delta}L_{z,30} - B(X_{30} - 1)}{{\Delta}L_{z,100} - B(X_{100} - 1)} & = \frac{Y - X_{30}}{Y - X_{100}}.
\end{align}
\end{subequations}
Let the left hand side of the last equation be equal to a constant $C$. Then the solution for $Y$ is
\begin{equation}
Y = \frac{-X_{30} + CX_{100}}{C-1}.
\end{equation}

For Reid Park and Tanque Verde, we find results for $Y$ (Table~\ref{priv-frac}) that are in strong agreement with one another. As measured at the two sites, the contribution to zenith luminance represented by ``private'' lighting decreases by about $1-Y\approx0.025$, or about 2.5\%, across midnight. For the other sites where we obtained measurements, the change in zenith luminance at midnight was too small relative to instrumental uncertainties to yield a meaningful estimate of $Y$.

Similarly, Equations 8a and 8b can be rearranged and, with Equation~\ref{total-bef}, used to solve for $L_{S,b}$:
\begin{subequations}
\begin{align}
L_{P,b} = \frac{{\Delta}L_{z,30}+B(1-Y)}{X_{30}-Y}\\
L_{P,b} = \frac{{\Delta}L_{z,100}+B(1-Y)}{X_{100}-Y}
\end{align}
\end{subequations}
Dividing the result for $L_{S,b}$ by $B$ then gives an independent, experimental determination of the fraction of total zenith sky brightness represented by street lighting before midnight. We find that this fraction varies from 1\% to 9\%, with mean values of 3\% at Reid Park and 2\% at Tanque Verde. These results are summarized in Table~\ref{priv-frac}.

\begin{table}
\centering
\begin{tabular}{ cccc }
 \hline
\textbf{Parameter} & \textbf{Reid Park} & \textbf{Tanque Verde} & \textbf{Units} \\
\hline
$B$ & $3.78\pm0.76$ & $0.93\pm0.02$ & mcd m$^{-2}$ \\
${\Delta}L_{z,30}$ & $-0.24\pm0.06$ & $-0.04\pm0.01$ & mcd m$^{-2}$ \\
${\Delta}L_{z,100}$ & $-0.08\pm0.03$ & $-0.02\pm0.01$ & mcd m$^{-2}$ \\
$C$ & $-4.67\pm0.01$ & $-4.62\pm0.01$ & mcd m$^{-2}$ \\
$Y$ & $0.974\pm0.001$ & $0.973\pm0.001$ & unitless \\
$L_{S,b}$/$B$ & $0.034\pm0.008$ & $0.018\pm0.007$ & unitless \\ 
 \end{tabular}
\caption{Summary of values used to calculate the fraction $Y$ by which the amount of zenith luminance attributable to lighting in Tucson other than known street lighting (``private lighting'') changed at midnight during the dimming experiment, and $L_{S,b}$/$B$, the fraction of zenith brightness at each site attributable to known street lighting before midnight.}
\label{priv-frac}
\end{table}

Lastly, we can use the above results to calculate the change in sky brightness at midnight attributable to private lighting alone: 
\begin{equation}
{\Delta}L_{z,P} = L_{P,a}-L_{P,b}.
\end{equation}
We presume for simplicity that this changes from night to night in some understandable way, such as a difference between week nights and weekend nights, when different temporal patterns in the use of light at night are expected. Further, we know that the condition of street lighting before midnight on any night is:
\begin{equation}
L_{s,b} = 0.9L_{s}(100),
\end{equation}
where the argument of $L_s$ means the zenith brightness contributed by the street lighting system if it were operated at 100\% of its nominal full power. If $L_{s}$(100) is known, then ${\Delta}L_{z,P}$ depends only on ${\Delta}L_z$.

Using Equations~\ref{eqn3a},~\ref{eqn3b} and~\ref{eqn4b}, we get:
\begin{subequations}
\begin{align}
L_{P,b}(30)[Y-1] & = {\Delta}L_z(30)+0.6L_{s}(100) 
\label{eqn17a}\\
L_{P,b}(100)[Y-1] & = {\Delta}L_z(100)-0.1L_{s}(100)
\label{eqn17b}
\end{align}
\end{subequations}
where the arguments of $L_{P,b}$ are the percentages on the two test night configurations (30\% and 100\%). If we assume that the contribution to the zenith brightness from private lighting before midnight is the same every night, then we can set the right-hand sides of Equations~\ref{eqn17a} and~\ref{eqn17b} equal to solve for $L_{s}$(100). Given our measured values for the change in zenith brightness on the 30\% and 100\% test nights from Reid Park, we find $L_{s}$(100) = ($0.22\pm0.08$) mcd m$^{-2}$. This implies that the change in private lighting alone across midnight, ${\Delta}L_{z,P}$ = ($-0.10\pm0.04$) mcd m$^{-2}$.

\section*{Acknowledgments}

The authors wish to thank the municipal government of the City of Tucson for participating in the experiment described here. They also recognize the valuable insights and comments of two anonymous reviewers whose suggestions helped sharpen the presentation. 

Funding: FK and MK acknowledge support from the Slovak Research and Development Agency under contract number APVV-18-0014. CK acknowledges funding from the Helmholtz Association Initiative and Networking Fund under grant ERC-RA-0031, as well as through the European Union's Horizon 2020 research and innovation programme ERA-PLANET, grant agreement no.~689443, via the GEOEssential project.

Conflicts of Interest: The authors declare that they hold no conflicts of interest in any aspect of the work presented herein.

Author roles: JB collected skyglow data, collated and analyzed skyglow measurements, and substantially wrote the manuscript text. FK and MK carried out SkyGlow Simulator runs and provided analysis of results. JS arranged for the dimming tests and provided technical assistance regarding the City of Tucson municipal street lighting system. GE, AMD, BF, and AG contributed skyglow observations from around Tucson during the tests. ST obtained spectra of skyglow over Tucson used to calibrate the magnitude-to-luminance conversion. CK conceived the idea of the test and provided critical feedback on the manuscript. All authors read and approved the manuscript.

\bibliography{barentine-jqsrt}

\end{document}